\documentclass[12pt,preprint]{aastex}

\newcommand{\kms}{km~s$^{-1}$} 
\newcommand{\kmssq}{km$^2$~s$^{-2}$} 
\newcommand{\formaldehyde}{H$_2$CO} 
\newcommand{\ethylcyanide}{CH$_3$CH$_2$CN}
\newcommand{\dimethylether}{CH$_3$OCH$_3$}
\newcommand{\methylcyanide}{CH$_3$CN}
\newcommand{\ttmethylcyanide}{CH$_3^{13}$CN}

\newcommand{\water}{H$_2$O} 
\newcommand{\hcccn}{HC$_3$N}  
\newcommand{\methanol}{CH$_3$OH}
\newcommand{\ammonia}{NH$_3$}

\newcommand{\lsun}{$L_\odot$}
\newcommand{\msun}{$M_\odot$}
\newcommand{\ngc}{NGC~6334}
\newcommand{\ngcf}{NGC~6334~F}

\newcommand{\ngce}{NGC~6334~E}
\newcommand{\ngcin}{NGC~6334~I(N)}
\newcommand{\mjb}{mJy~beam$^{-1}$}
\newcommand{\jb}{Jy~beam$^{-1}$} 
\newcommand{\mujb}{$\mu$Jy~beam$^{-1}$}
\newcommand{\pcsq}{pc$^{2}$}

\newcommand{\percms}{cm$^{-2}$}
\newcommand{\percmc}{cm$^{-3}$}
\newcommand{\pcc}{pc$^{3}$}
\newcommand{\ppcc}{pc$^{-3}$}

\begin{document}

\shortauthors{Hunter et al.}

\shorttitle{The Massive Protocluster \ngcin\ }

\title{Subarcsecond Imaging of the \ngcin\/ Protocluster:\\ Two Dozen Compact Sources
 and a Massive Disk Candidate}

\author{T. R. Hunter\altaffilmark{1}, C. L. Brogan\altaffilmark{1}, 
  C. J. Cyganowski\altaffilmark{2,3}, K. H. Young\altaffilmark{2}
}
 
\email{thunter@nrao.edu}

\altaffiltext{1}{NRAO, 520 Edgemont Rd, Charlottesville, VA 22903} 
\altaffiltext{2}{Harvard-Smithsonian Center for Astrophysics, Cambridge, MA 02138}  
\altaffiltext{3}{SUPA, School of Physics and Astronomy, University of St. 
Andrews, North Haugh, St. Andrews KY16 9SS, UK}  

\slugcomment{accepted for publication in {\it The Astrophysical Journal}}

\begin{abstract}

Using the SMA and VLA, we have imaged the massive protocluster
\ngcin\/ at high angular resolution ($0.5'' \sim 650$~AU) from
6~cm to 0.87~mm, detecting 18 new compact continuum sources.  Three of
the new sources are coincident with previously-identified \water\/
masers.  Together with the previously-known sources, these data bring
the number of likely protocluster members to 25 for a protostellar
density of $\sim 700$ \ppcc.  { Our preliminary measurement of} the
$Q$-parameter of the minimum spanning tree is 0.82 -- close to the
value for a uniform volume distribution.  All of the (nine) sources
with detections at multiple frequencies have SEDs consistent with dust
emission, and two (SMA~1b and SMA~4) also have long wavelength
emission consistent with a central hypercompact HII region. Thermal
spectral line emission, including CH$_3$CN, is detected in six
sources: LTE model fitting of CH$_3$CN ($J$=12--11) yields
temperatures of 72--373~K, confirming the presence of multiple hot
cores.  The fitted LSR velocities range from $-3.3$ to $-7.0$~\kms,
with an {unbiased} mean square deviation of { 2.05~\kms},
implying a protocluster dynamical mass of {410$\pm$260~\msun\/}.
From analysis of a wide range of hot core molecules, the kinematics of
SMA~1b are consistent with a rotating, infalling Keplerian disk of
diameter 800~AU and enclosed mass of 10-30~\msun\/ that is
perpendicular (within $1\arcdeg$) to the large-scale bipolar outflow
axis.  A companion to SMA~1b at a projected separation of
$0.45\arcsec\/$ (590~AU; SMA~1d), which shows no evidence of spectral
line emission, is also confirmed.  Finally, we detect one 218.4400~GHz
and several 229.7588~GHz Class-I \methanol\/ masers.



\end{abstract}
 

\section{Introduction}

Massive star formation is a phenomenon of fundamental importance in
astrophysics yet a detailed understanding of it has been elusive
\citep{Zinnecker07}. {The present state of theory and numerical
simulation research fall into two major categories of processes:
``core accretion'' and ``competitive accretion'', as recently reviewed
by \citet{Tan14}.  In the core accretion scenario, massive stars (like
low-mass stars) are formed via the collapse of self-gravitating,
centrally condensed cores.  These cores are discrete structures within
the larger-scale molecular cloud, and each core constitutes the mass
reservoir for a single star or small multiple system.  As a result,
core mass maps directly to stellar mass, and the core mass function
(CMF) maps to the stellar initial mass function (IMF) \citep[e.g.][and
  references therein]{Myers13, Krumholz09,McKee03}.  In contrast,
competitive accretion models are intrinsically models of star cluster
formation.  Fragmentation in a cluster-scale gas clump produces many
low-mass protostars, which then accrete (competitively) from the
large-scale gas reservoir.  In this picture, massive stars \emph{must}
form in a cluster environment, and massive stars and their surrounding
clusters must form simultaneously \citep[e.g.][and references
  therein]{Bonnell11,Smith09}.}

There are several significant observational constraints that these
theories must face.  First, there is mass segregation -- the fact that
the most massive members of young clusters are concentrated in the
center \citep[e.g.][]{Kirk11}.  It remains unclear whether this
property is primordial or a result of dynamical evolution.  $N$-body
simulations of star clusters have shown that it is common to form
compact groupings of massive stars (i.e. Trapezium-like systems) near
the center of the cluster in as little as one free-fall time,
particularly when there is initial substructure \citep{Allison11}.
Observational studies of young clusters caught early in the act of
forming are critical to address the origin of mass segregation.
Second, there is the correlation between the mass of the most massive
star and the number of cluster members \citep{Hillenbrand95,Testi99}.
An ensuing question is whether the high-mass and low-mass members of
clusters form at the same time.  Perhaps the only way to answer this
question is to search for actively forming low-mass protostars amidst
their high-mass counterparts.  Toward this end, the past decade of
advances in (sub)millimeter interferometry has enabled the
identification of proto-Trapezium-like systems, termed
``protoclusters'' \citep[e.g.][]{Hunter06,Beuther07a,Rodon08}, in
which four or more compact millimeter continuum sources exist within a
projected diameter of 10000~AU.  In many cases, the protocluster
members present a striking diversity in star formation indicators --
including the presence or absence of \water\/ masers, hot core line
emission, and free-free emission -- providing strong evidence that
closely neighboring objects can exist in different evolutionary stages
of formation
\citep{Zhang07,Cyganowski07,Brogan08,Brogan11,Zinchenko12}.  At last
count of the literature, half of these massive cores studied at
$\lesssim 1000$~AU resolution were resolved into four or more
millimeter sources \citep{Palau13}.  These results emphasize the
possibility of interaction between massive protostars, as well as
their probable impact on low-mass protostars forming in their midst.
Furthermore, given the deeply embedded nature of protoclusters,
sensitive high dynamic range millimeter and centimeter imaging will be
needed to obtain an accurate census of their low-mass membership.

A third observational constraint on theory is the tentative yet
growing evidence for massive Keplerian accretion disks around massive
protostars.  Discovered through subarcsecond angular resolution
imaging, there is a steadily increasing number of massive disk
candidates around central stars of varying mass $M_*$.  Examples of
disk candidates with an enclosed mass of $M_* \approx 7-10$~\msun\/
include IRAS~20126+4104, IRAS~18360-0537, and Orion KL Source I
\citep{Cesaroni05,Xu12,Qiu12,Hirota14}.  Larger scale structures
(toroids) of diameter several thousand AU have also been reported
which may encompass either an O-type protostar or a cluster of massive
protostars \citep[e.g.][]{Beltran11,Zapata10}.  There are also cases
of apparent sub-Keplerian rotation \citep[AFGL2591-VLA3:][]{Wang12} as
well as of a lack of Keplerian rotation signatures on 500~AU scales
\citep[NGC7538~IRS1:][]{Beuther13}.  Most of the candidate disks show a
strong bipolar molecular outflow perpendicular to the disk plane,
analogous to low-mass protostellar disk/outflow systems.  It is
important to note that the presence of such disks does not immediately
favor either the core accretion or competitive accretion model, as
they are expected to exist under both scenarios.  However, as our
knowledge of the massive disk population grows, both theories are
certain to face new challenges.

To further explore the protostellar population of massive
protoclusters, we have been pursuing detailed observations of the
nearby examples in \ngc\/, a region containing multiple sites of high
mass star formation \citep{Straw89a,Persi08,Russeil10}. {A recent deep
near- and mid-infrared survey revealed over 2200 young stellar object
(YSO) candidates, and subsequent estimates of the star formation rate
suggest that it may be undergoing a ``mini-starburst'' event
\citep{Willis13}}.  At the northeastern end of the region, the
deeply-embedded source ``I(N)'' was first identified at 1~mm by
\citet{Cheung78} and later detected at 400~\micron\ by
\citet{Gezari82}, who estimated a size of 50\arcsec.  Single-dish
observations of high velocity SiO emission indicated significant
outflow activity at this location \citep{Megeath99}.  Our initial SMA
observations of \ngcin\ at $\sim2$\arcsec\/ resolution resolved a
Trapezium-like protocluster of seven compact millimeter continuum
sources within a projected diameter of 0.1~pc
\citep{Hunter06,Brogan09}.  As revealed by the millimeter spectral
line data \citep{Brogan09}, the brightest three continuum sources
(SMA~1, SMA~2, and SMA~4) are the origin of the hot core line emission
seen in the many single-dish molecular line observations of this
source \citep{Kuiper95,Thorwirth03,Thorwirth07,Walsh10,Kalinina10}.
Hot \ammonia\/ was resolved by 1.5\arcsec\/ resolution Australia
Telescope Compact Array (ATCA) observations of the (5,5) and (6,6)
lines, which peak toward SMA~1 \citep{Beuther07b}.  The profile of the
(6,6) line showed a double peak separated by $\sim$4~\kms, which was
interpreted as possibly tracing a rotating circum-protostellar disk
\citep{Beuther07b}. With similar angular resolution, \citet{Brogan09}
detected a comparable velocity gradient toward SMA~1 in a few other
hot core molecules.  However, Karl G. Jansky Very Large Array (VLA)
7~mm continuum observations with a 0.5\arcsec\/ beam resolved SMA~1
into multiple components \citep{Brogan09,Rodriguez07}, making the
larger-scale velocity gradient difficult to interpret.

\begin{figure}
\epsscale{0.8}
\plotone{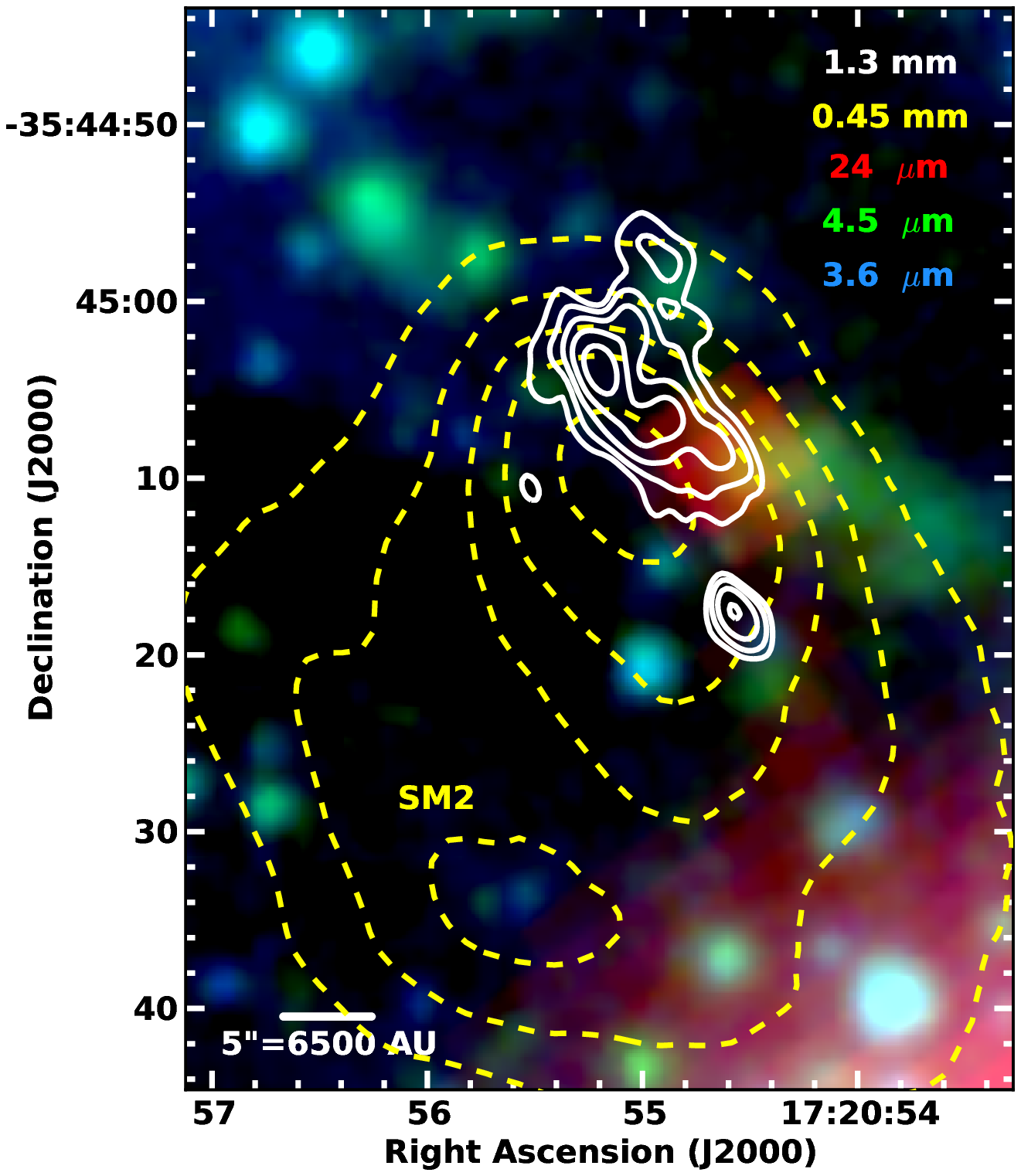} 
\caption{Overview of the NGC6334 I(N) region in the mid-infrared.  The
  three-color image was generated from the {\it Spitzer} MIPSGAL and
  GLIMPSE survey images at 24, 4.5, and 3.6~\micron\/ \citep{Benjamin03,
    Churchwell09,Carey09}. For reference the SMA compact + extended (COMP+EXT)
  configuration 1.3~mm continuum with $2.2\arcsec\times 1.3\arcsec$
  resolution from \citet{Brogan09} is shown in white contours
  (levels=40, 80, 160, 320, 640~\mjb). The dashed yellow contours show
  the JCMT 0.45~mm emission ($14''$ resolution) from \citet{Sandell00}
  (levels: 60, 80, 100, 120, 160~\jb).  The labeled 0.45~mm contour
  indicates the position of the single-dish source SM2 reported by
  \citet{Sandell00}.
\label{fig0}}
\end{figure}

The \ngcin\/ millimeter protocluster is embedded in a region that is
remarkably dim in the mid-infrared {\it Spitzer} images, and has the
characteristics of an Infrared Dark Cloud (IRDC).  To provide an
overview, Figure~\ref{fig0} reprises the millimeter and infrared
continuum imaging results from \citet{Brogan09}. Despite the
millimeter multiplicity, we identified only some extended
4.5~\micron\/ emission associated with the bipolar outflows from
SMA~1, 4 and 6, and a single 24~\micron\/ point source near SMA~4
\citep{Brogan09,Hunter06}.  In contrast, our VLA detection of copious
amounts of 44~GHz Class I \methanol\ maser emission to the southeast
of the compact millimeter continuum sources is indicative of outflow
activity \citep{Cyganowski09,Kurtz04,Voronkov14}, specifically in the
area surrounding the single-dish (sub)millimeter continuum source SM2
\citep{Sandell00}.  Similarly, our VLA \water\ maser observations
revealed 11 locations of emission, only 8 of which were associated
with the known compact millimeter continuum sources, suggesting the
presence of additional young stellar objects \citep{Brogan09}.



In this paper, we present new sub-arcsecond SMA 1.3~mm, 0.87~mm and
VLA 6~cm imaging that indeed reveals significant further multiplicity
in this protocluster, as well as the detailed kinematics of the hot
cores and complex spectral energy distributions.  The details of the
observations are summarized in \S~\ref{obs}, while \S~3 and \S~4
present our key results and discussion, respectively. For the distance
to \ngcin\/, we adopt 1.3 kpc based on recent \water\/ maser parallax
studies: $1.34^{+0.15}_{-0.12}$~kpc \citep{Reid14} and
$1.26^{+0.33}_{-0.21}$~kpc \citep{Chibueze14}. In the past, the most
commonly used value was 1.7 kpc from photometric estimates for the
\ngc\/ region \citep{Neckel78,Pinheiro10}, implying a reduction by a
factor of 1.7 for derived quantities based on the distance squared,
such as mass and luminosity.  For example, the total luminosity of
\ngcin\/ as measured by \citet{Sandell00} is now $1.0 \times
10^3$~\lsun\/ with this revised distance.


\section{Observations}
\label{obs}

The details of the SMA\footnote{The Submillimeter Array (SMA) is a
  collaborative project between the Smithsonian Astrophysical
  Observatory and the Academia Sinica Institute of Astronomy \&
  Astrophysics of Taiwan.} 1.3~mm and 0.87~mm observations and the
  NRAO\footnote{The National Radio Astronomy Observatory is a facility
  of the National Science Foundation operated under cooperative
  agreement by Associated Universities, Inc.} VLA 6~cm observations
  are summarized in Table~\ref{smaobs}.  The very extended
  (henceforth, VEX) 1.3~mm SMA data were calibrated in MIRIAD
  \citep{Sault95}, then exported to CASA \citep{McMullin07} where the
  continuum was subtracted to create a line-only VEX dataset. The
  continuum is then composed of the line-free portions of the
  dataset. Self-calibration was performed on the continuum data, and
  solutions were transferred to the line data.  Line cubes were
  generated with a channel spacing of 1.1~\kms\/ and robust weighting
  of 1.0, with the exception of the 218.44 and 229.76 GHz Class I
  methanol maser transitions which were imaged with robust weighting
  of 0.0.  The flux calibration is based on Titan, Ceres, and SMA flux
  monitoring of the observed quasars and is estimated to be accurate
  to within 20\%. The same procedure was also followed for the VEX
  0.87~mm data. Unfortunately, the observing conditions for the
  0.87~mm data were significantly worse than for the 1.3~mm data in
  terms of higher winds and more variable opacity, leading to greater
  phase instability.  As a result, the 0.87~mm spectral line cubes are
  too noisy to be useful.


Two 1.3~mm continuum images were constructed: (1) to maximize the
continuum sensitivity and minimize artifacts from resolved out
structure, the VEX 1.3~mm continuum data were combined with the
extended configuration (henceforth, EXT) continuum data presented in
\citet{Hunter06} and \citet{Brogan09} and imaged with robust weighting
of 0.5. Note that the EXT 1.3~mm data only cover a portion of the
spectral coverage of the VEX data so no such combination was possible
for the line data. 
The relative weight of the individual visibilities between the two
configurations is such that the angular resolution of the combination
is similar to that of the VEX data alone.  This ``EXT+VEX'' continuum
image was used to identify and characterize the 1.3~mm dust continuum
sources; it is not sensitive to smooth structures larger than about
$9''$.  (2) In an effort to better match the uv-coverage of the VEX
0.87~mm data, the line-free portions of the VEX 1.3~mm data were
imaged with uv-spacings $> 90$k$\lambda$ and robust=0 weighting. This
image is henceforth termed the ``VEX-UV'' 1.3~mm continuum image. The
0.87~mm continuum image was also created two ways, both using robust
weighting of 1.0: (1) a version using all of the data to obtain good
angular resolution and sensitivity; and (2) a version with a
300k$\lambda$ uv-taper to better match the VEX 1.3~mm data, which was
then subsequently convolved to the same resolution as the VEX-UV
1.3~mm continuum image. This uv-tapered and convolved image was used
to determine the 0.87~mm dust continuum properties. The 0.87~mm images
are not sensitive to smooth structures larger than about $3''$. All
measurements from all images were obtained from versions that had been
corrected for the primary beam response.

The VLA data were calibrated in CASA with scripts based on the VLA
pipeline\footnote{\url{https://science.nrao.edu/facilities/vla/data-processing/pipeline}}.
Imaging and self-calibration were performed manually in CASA.  The
flux calibration accuracy is estimated to be 10\%.  Due to the large
primary beam (nearly $10'$), the VLA 6~cm data include the bright
ultracompact HII region \ngcf\/ as well as the 45\arcsec\/-diameter
HII region \ngce\/ \citep{Rodriguez03}, which are both located south
of the field of interest.  We thus imaged a large field ($10^7$
pixels) containing these objects in order to minimize the confusion
toward \ngcin. {Comparing the new 6~cm JVLA data to the 1.3~cm and 7~mm
VLA data presented in \citet[][see also Rodriguez et
  al.\ 2007]{Brogan09}, we found a $\sim 0.4\arcsec$ disagreement in
the astrometry between the old and new data. The most likely culprit
was the phase calibrator used in the older VLA observations J1720-358
(B1717-358). We alerted both VLA and ALMA to our suspicion, which led
to an ALMA calibration observation of J1720-358 using the nearby VLBA
calibrator J1717-3342 \citep{Petrov06} at Band 3 (92 GHz) as the
reference source with 24 antennas (Oct. 29, 2013, execution blocks:
uid://A002/X70c186/X12, X25, and X4a). As a result of these
observations, the ALMA calibrator database has been updated with a
revised position of 17:20:21.798$\pm$0.030\arcsec,
-35:52:48.128$\pm$0.010\arcsec\/ (a correction of 0.43\arcsec\/, as of
this writing the VLA calibrator database has not yet been updated).} We
have used this new information to correct the astrometry of the 1.3~cm
and 7~mm images used in this paper. We note that our \water\/ maser
observations used J1717-3342 as the phase calibrator \citep{Brogan09}
and thus do not require correction.



\section{Results} 

\subsection{Continuum emission}

\subsubsection{1.3~mm continuum}
\label{annulus}

The 1.3~mm EXT+VEX image of the continuum emission is shown in
Figure~\ref{fig1}.  All of the seven previously-identified millimeter
sources \citep{Brogan09} are detected, with the exception of one of
the fainter, more extended sources (SMA~7), whose emission is
apparently resolved out by the longer baselines employed in these
observations.  We detect many new compact sources in the surrounding
field, primarily because the rms noise achieved is nearly 4 times
better than \citet{Brogan09}, and the higher angular resolution leads
to a lower confusion limit.  Due to the limited size of the SMA
primary beam, the rms noise in the image corrected for primary beam
response increases as a function of angular radius from the phase
center.  Therefore, we have established our detection threshold
corresponding to $N\sigma(a)$ where $\sigma(a)$ is the rms noise
measured in an annulus $a$.  We defined four annuli corresponding to
the following levels in the CASA sensitivity image (which for a single
pointing observation is simply the image of the primary beam
response): 80\%--100\%, 60\%--80\%, 40\%--60\%, and 20\%--40\%, for
which we measured $\sigma(a)$ values of 2.2, 2.5, 3.4 and 5.4~\mjb.
The area of the image above the 20\% sensitivity level corresponds to
5500 times the area of the synthesized beam.  Therefore, we chose
%
$N=4.5$ because the expected number of false $+4.5\sigma$ ``peaks'' is
only 0.019.  (For $4\sigma$, the value is 0.17 which we considered to
be too close to 1 false positive.)  With our $4.5\sigma$ criterion, we
identify 24 sources, 16 of which are new (Table~\ref{contsources}),
and two of which (SMA~20 and 18) correspond to two of the brightest
\water\ maser components \citep[C1 and C2, respectively;][]{Brogan09}.
All of the new sources, starting with SMA~8, are numbered in order of
increasing right ascension.  The majority of new sources are found in
the southern half of the SMA primary beam, consistent with the area of
extended dust emission seen in the single dish 0.45~mm image (see
Figs~1 and 2).  SMA~18 coincides with the single-dish source SM2 to
within the single-dish position uncertainty \citep{Sandell00}.  We
used the CASA imfit task to fit each source with a two-dimensional
Gaussian to find the peak position and integrated flux density and
attempt to find the deconvolved size.  For 9 of the 24 sources, the
fitted size was well constrained in both the major and minor axes, and
we can compute the brightness temperature (a lower limit to the
physical temperature).  For 11 sources, the minor axis was not well
constrained, and the geometric mean of the major and minor axes of the
fitted Gaussian was less than that of the synthesized beam
(0.5\arcsec).  Four sources were consistent with a point source in
both axes.  For the latter two cases, we assign 0.5\arcsec\/ as an
upper limit to the size and use this to compute a lower limit to the
brightness temperature.  With the exception of the 24~\micron\/ source
near SMA~4, which likely traces hot dust in the walls of the cavity
formed by the outflow from that source \citep[e.g.][]{DeBuizer05},
none of the millimeter continuum sources have counterparts in the
mid-infrared {\it Spitzer} images \citep{Brogan09}.

\subsubsection{6~cm to 7~mm continuum and revised astrometry}

The VLA 6~cm image (Figure~\ref{fig1}) shows two previously known 6~cm
point sources, associated with SMA~1 and SMA~4 \citep{Carral02}, plus
two new sources that are not associated with 1.3~mm emission.  All
four sources are consistent with unresolved compact sources with sizes
$<0.3$\arcsec.  Their positions and flux densities are listed in
Table~\ref{cmcontsources}.  One of the new sources coincides within
0.054\arcsec\/ (70~AU) with the centroid of \water\/ maser component
C4 \citep{Brogan09}, hence we call it \water\/-C4.  The other new
source is located $\approx$30\arcsec\/ (0.2~pc) south of
SMA~1. Because it is not coincident with any maser or 1.3~mm source,
we call this source VLA~2 in order to distinguish it from both the
1.3~cm source VLA-K1 reported by \citet{Rodriguez07} (outside the SMA
field of view) and the 7~mm source VLA~3 reported by \citet{Brogan09}
(which was named for its proximity to SMA~3 and yet independent
nature).

The central portion of the \ngcin\/ protocluster is shown in
Figure~\ref{sma1zoom}.  The 1.3~mm emission from SMA~1 has been
clearly resolved into three components (a, b+d, and c), which are
composed of the four sources previously identified at 7~mm.  In
addition, the east/west extension of SMA~1b at 1.3~mm is consistent
with the presence of the fourth 7~mm component (SMA~1d).  Both SMA~1b
and SMA~1d are detected at 1.3~cm and 7~mm while only SMA~1b is
detected at 6~cm.  Revised positions of the 1.3~cm and 7~mm
counterparts based on the corrected astrometry of those images are
given in Table~\ref{cmpositions}.  With the new astrometry, there is
now very good agreement between the VLA and SMA continuum contours for
the primary millimeter sources (SMA~1b+d, 4, and 6).  The angular
separation between SMA~1b and 1d in the 7~mm image is 0.45\arcsec\/
(590~AU) at a position angle of -67\arcdeg\/ (east of north).  The
area surrounding SMA~6 is shown in Figure~\ref{sma6zoom}.  The peaks
of the 7~mm, 1.3~mm and 0.87~mm emission are in good agreement.  There
is an extended ridge of 1.3~mm emission to the southwest, part of
which may arise from a distinct source.  However, because it is not
clearly separated from SMA~6 and there is no compact counterpart at
any other wavelength, we have not identified this ridge as a separate
object.

\subsubsection{0.87~mm continuum}
\label{missingflux}

A further zoomed view of the SMA~1b+d region is shown in
Figure~\ref{sma1bd}a,b in 1.3~mm and 0.87~mm continuum, along with the
\water\/ maser positions {observed with a beam of $0\farcs 79 \times
0\farcs 25$ at P.A.=+7\arcdeg} \citep{Brogan09} and the first moment of
the \methylcyanide\/ $J$=12--11, $K$=7 transition.  With higher
resolution than the 1.3~mm VEX image, the 0.87~mm image clearly
indicates that SMA~1d is a source of submillimeter emission separate
from SMA~1b.  Further evidence that these are distinct sources comes
from the \water\/ maser positions: the vast majority are coincident
with SMA~1b while none are seen toward SMA~1d \citep[see
  also][]{Chibueze14}.  The position angle of the spatial extent of
the \water\/ masers is somewhat inclined to that of the large-scale
bipolar outflow axis measured in SiO 5--4 \citep{Brogan09}.  Also, the
kinematics are complicated, and reminiscent of the \water\/ maser
system of radio source I in Orion~BN/KL \citep[see
  e.g.][]{Greenhill13} in terms of spatial extent ($\sim$500~AU) and
the apparent overlap of widely disparate velocities. Similar to the
\water\/ emission, the \methylcyanide\/ emission is centered on SMA~1b
(see \S~\ref{hotcore}), with no measurable emission coming from
SMA~1d.

In an attempt to apportion the 1.3~mm continuum flux density between
SMA~1b and 1d, we fit a single two-dimensional Gaussian to SMA~1b in
the 1.3~mm VEX-UV continuum image.  The residual image showed
a weak source of unresolved emission coincident with the 7~mm and
1.3~cm source SMA~1d. To estimate the flux density of this source, we
fit the residual image with a single Gaussian, which yielded a point
source of $44 \pm\/ 8$~mJy at the J2000 position:
17:20:55.23$\pm$0.07\arcsec, -35:45:04.14 $\pm$ 0.07\arcsec.  The
angular separation of this position from SMA~1b is
0.56$\pm$0.04\arcsec\/ at position angle -69$\pm$10\arcdeg, in good
agreement with the 7~mm separation between SMA~1b and 1d.  Finally, we
estimate the flux density of SMA~1b alone to be the joint flux density
of SMA~1b+d from Table~\ref{contsources} minus the value for SMA~1d,
or $1.02\pm\/0.08$~Jy.  A similar procedure was then performed on the
0.87~mm image, yielding a flux density of $100\pm\/25$~mJy for SMA~1d.

Eight of the 1.3~mm sources are detected in the uv-tapered 0.87~mm
image. Note that 10 of the 16 non-detected sources (including VLA~2)
lie beyond the one-third sensitivity radius of the primary beam, where
the $4.5\sigma$ limit is $>0.12$~\jb.
Using this image, the fitted positions, flux densities, and sizes are
listed in Table~\ref{contsourcesBand7}.  We caution that the
uv-tapered 0.87~mm image is still not sensitive to the largest angular
scales that the 1.3~mm EXT+VEX image contains.  Therefore, the 0.87~mm
flux densities should be taken as lower limits for those sources
having 1.3~mm fitted sizes significantly larger than $0.5\arcsec$,
particularly those located in the complicated central cluster.  For
the 1.3~mm sources smaller than $0.5\arcsec$ and located away from the
central cluster (including SMA~13 and SMA~15), the 0.87~mm
measurements are unlikely to suffer from missing flux and may be
considered to be accurate.

\subsection{Hot core line emission} 
\label{hotcore}
 
In our SMA 1.3~mm data cubes, SMA~1b, SMA~2 and SMA~4 all show copious
hot core line emission, as do SMA~6, SMA~15, and SMA~18 to a lesser
extent.  At the current sensitivity, 1.3~mm spectral line emission is
not detected toward any of the other cores. The spectral line emission
is dominated by spatially compact emission from complex
molecules. Indeed, most of the more abundant species that exhibited
outflow emission (e.g. CO, $^{13}$CO, SiO, DCN, etc.) in the lower
resolution SMA data presented in \citet{Brogan09} are mostly resolved
out in the very extended configuration data. As a result, the channels
with these molecules have significantly higher noise (by factors of
3-10) due to imaging artifacts caused by the lack of short spacing
information. From the 8 GHz of available 1.3 mm spectral bandwidth we
identified for detailed analysis 12 transitions from 9 different
chemical species that are representative of the range of line emission
morphologies, kinematics, and line excitation temperatures detected in
these data, and furthermore do not suffer significant line blending
(see Table~\ref{linelist} for more details).  First moment images of
these transitions toward the central region of the cluster are shown
in Figure~\ref{moments}, where the field of view is the same as
Figure~\ref{sma1zoom}.  The parent cube was made with Briggs weighting
and a robust parameter of 1.0.  Details of the transitions shown are
provided in Table~\ref{linelist}, including their rest frequencies,
excitation energies, and integrated intensities.  The first striking
feature in the images is the velocity gradient across the central
0.8\arcsec\/ (1000~AU) of SMA~1b, which is seen consistently in all
transitions.  The orientation and magnitude of the velocity gradient
is comparable to that seen at $\sim$2\arcsec\/ angular resolution
\citep{Brogan09}.  In the lower energy lines such as HC$_3$N and OCS,
the emission and its accompanying gradient also extend somewhat to the
northeast toward SMA~1c. A zoomed view of the \methylcyanide\/ $K$=7
transition is shown in Figure~\ref{sma1bd}b. The magnitude of the
gradient ranges from 3 to 5 \kms/arcsec, {with the largest values
  seen in the lines of \methanol\/ and CH$_3$OCHO}.
The position angle of the gradient was measured in each transition by
viewing the data cube in CASA, marking the RA/Dec centroid of the
emission in the outer velocity channels, and computing the
corresponding slope.  The median and standard deviation of the
position angle taken over all the transitions is -52$\pm$5\arcdeg.

SMA~2 and SMA~4 also show compact emission, {but in only 9 or 10
  of the 12 transitions, respectively.}  The differences in detections
indicate differences in both chemistry (HC$_3$N being detected in
SMA~4 but not SMA~2) and excitation conditions (SMA~2 being more
extended in the lower energy transitions).  Also, the central velocity
of the emission in SMA~1b differs by several \kms\/ from the emission
seen toward SMA~2 and SMA~4, indicating a significant velocity
dispersion in this protocluster.  To better quantify the
characteristics of the line emission and estimate the gas physical
conditions, we extracted the $J$=12--11 \methylcyanide\/ spectra at
the peak positions of all the continuum sources in
Table~\ref{contsources}.  Only 6 sources show detectable emission; in
particular, there is no sign of emission from SMA~1c or 1d.  We used
the CASSIS\footnote{CASSIS has been developed by CESR-UPS/CNRS
  (\url{http://cassis.cesr.fr}).} package to perform a local
thermodynamic equilibrium (LTE) model fit to the \methylcyanide\ and
\ttmethylcyanide\ ladders, which provide a good measurement of gas
temperature in hot cores \citep[e.g.][]{Pankonin01, Araya05}.  There
were five free parameters in the model: temperature, velocity, line
width, column density, and diameter.  Based on the galactocentric
distance of \ngc, we assume a $^{12}$C:$^{13}$C ratio of 58
\citep{Milam05}.  {We used the Markov Chain Monte Carlo (MCMC)
  $\chi^2$ minimization option, which is far more CPU time-efficient
  and tractable than a uniform grid search.  Unlike a grid search, the
  MCMC search begins with an initial guess and a range for each
  parameter, and makes random steps in each parameter in order to
  explore the range.  The amplitude of the random step size is
  initially large but decreases during the first stage of the
  computation as the space is explored.  In the latter stage, smaller
  steps are used to refine the best chi-squared potential.  In order
  to establish the parameter space for each source and the initial
  guess, we ran initial executions with very broad ranges in order to
  encompass plausible values.  Based on these results, we narrowed the
  ranges somewhat. The typical search parameters were ranges of
  $\pm60$~K for cool sources and $\pm100$~K for warm sources in
  temperature, $\pm 2$\kms\ in linewidth, $\pm 3$\kms\ in velocity,
  and a factor of 10-50 in column density (all centered on the initial
  guess), and a range of 0.1\arcsec\/-0.5\arcsec\/ for angular
  diameter.}  With these ranges, we ran several executions of MCMC
each with a different value of the automatic step size reduction
factor (called reducePhysicalParam), thereby achieving a variety of
acceptance rates in the range of 0.2 to 0.4 after 2000
iterations. {The cutoff parameter, which determines when the step
  size reduction factor becomes fixed, was set to half the number of
  iterations.}

The best fit values reported in Table~\ref{temperatures} are taken
from the execution that achieved an optimal acceptance rate of
$\approx$0.25. To obtain realistic uncertainties for the parameters,
we take the standard deviation from this single execution and add (in
quadrature) half the total spread in the individual best fit model
values.  The best-fit model spectra are overlaid on the observed
spectra in Fig.~\ref{linefits}.  In the case of SMA~4, the line shapes
are clearly non-Gaussian and a single component fit yields an
abnormally high line width of $>8$~\kms. For this source, we also
performed a two component fit, which resulted in smaller, more
reasonable line widths from two spatially unresolved components that
have rather different temperatures and velocities (by 6~\kms).  This
result may indicate the presence of an unresolved binary hot core
system.  For SMA~1b, we also found that two temperature components
produced a better fit to the low and high-$K$ lines of the spectrum,
as is typical of many hot cores
\citep[e.g.][]{Hernandez14,Cyganowski11}. {In this case, the
  search range for the cooler component was 60-150~K, and for the hot
  component was 150-600~K.}  The best-fit temperature of the hot
component of SMA~1b being in excess of 300~K is consistent with the
detection of the $K$=10 component with $E_{\rm lower} = 771$~K.  The
final column in Table~\ref{temperatures} provides an
order-of-magnitude estimate of the volume density of H$_2$ that was
calculated from the fitted column density assuming spherical geometry
and an assumed \methylcyanide\/ abundance of $1\times 10^{-8}$.  This
value is intermediate between the extrema of recent measurements of
the \methylcyanide\/ abundance in hot cores, which range from $1\times
10^{-7}$ to $1\times 10^{-9}$ \citep[see
  e.g.][]{Remijan04,Hernandez14}.  A comparison of the
\methylcyanide\/ column density to the dust column density is
discussed further in \S~\ref{dustcalc}.


\subsection{Maser Action in two 1.3mm \methanol\/ lines} 

\label{masersection}

Probable maser emission in the Class I 229.75881~GHz
CH$_{3}$OH~(8$_{\rm -1,8}$--7$_{\rm 0,7}$)~E transition is a
conspicuous feature of recent SMA observations of massive protostellar
outflows \citep[e.g.][and references therein]{Cyganowski12,
  Cyganowski11}. In these outflows, the 229 GHz emission is generally
co-located (spatially and spectrally) with lower frequency Class I
masers.  In \ngcin\/, we observe strong features in both the 229~GHz
transition and the Class I 218.44005 GHz (4$_{2,2}$--3$_{1,2}$)~E
transition at the position of the brightest 44 GHz maser \citep[spot
  55 of ][]{Brogan09}, which also coincides in velocity and position
with the $-3$~\kms\/ 24.9~GHz \methanol\/ maser
\citep{Beuther05,Menten89}.  Although interferometric observations of
the 229~GHz line have been reported previously
\citep[e.g.][]{Fish11,Cyganowski11,Cyganowski12}, our observations are the
first with sufficient angular resolution to test unequivocally whether
the brightness temperature ($T_B$) exceeds the upper energy level of
the transition ($E_{upper}$).  The peak values of $T_B$ in the 229~GHz
and 218~GHz lines (3100~K and 270~K, respectively) are both
significantly higher than the $E_{upper}$ of these transitions (89~K
and 45~K, respectively), indicative of maser emission.  Strong,
point-like emission also appears at several other positions in these
lines, but primarily in the 229~GHz line
(Figure~\ref{methanolmasers}).  We note that the 229~GHz line is part
of the 36~GHz maser series, while the 218~GHz line is part of the
25~GHz maser series \citep{Voronkov12}.  All positions with peak
intensity greater than 0.55~\jb\/ ($\sim 55$~K) are listed in
Table~\ref{masertable}; this threshold was chosen because it excludes
all of the emission seen toward the hot cores. Seven of the nine
components are within 0.2\arcsec\/ of a 44~GHz maser \citep{Brogan09}
and/or within 0.33\arcsec\/ of a 24.9~GHz \methanol\/ maser
\citep{Beuther05}.  All nine have brightness temperatures greater than
the energy level of the transition.  In both transitions, fainter
emission appears toward the continuum sources SMA~1, SMA~2 and SMA~4.
This emission is likely to be thermal because the brightness
temperatures are below 50~K.  Finally, there are two candidate maser
positions in the 218~GHz line which do not meet our peak intensity
threshold, but do have brightness temperatures higher than 50~K and
are not associated with a continuum source.  One corresponds in
position and velocity to component 4 of the 229~GHz line, and the
other lies at 17:20:53.781, -35:45:13.58 (J2000) with an LSR velocity
of $-$1.3~\kms.

\section{Discussion}

\subsection{Nature of the continuum sources}

Nine of the continuum sources are detected at more than one
wavelength.  To explore the nature of these objects, we have
constructed the spectral energy distributions (SEDs) in
Figure~\ref{seds}.  Two of the objects, SMA~1b and SMA~4, are detected
longward of 1.3~cm, and their SEDs are well-described
by a combination of free-free emission plus dust emission.  The
other seven sources all show a steeply rising spectral index in the
submillimeter.  Note, because the existing 3~mm observations
\citep{Beuther08} have more than four times poorer angular resolution
than the other wavelengths presented here, it is not possible to
separate the contributions to the 3~mm flux density for the individual
components of SMA~1.  Additionally, the 3~mm flux densities for SMA~4
and SMA~6 are regarded as upper limits for the size scales considered
here.  Further details on individual sources are described below
{along with descriptions of the modelling procedure}.

\subsubsection{SMA 1b and SMA 4}


With the addition of the very sensitive 6~cm data and the improved
astrometry at 1.3~cm and 7~mm compared to \citet{Brogan09}, it is now
possible and appropriate to do somewhat more sophisticated modeling of
the free-free emission from SMA~1b and SMA~4, since the millimeter and
centimeter emission have been shown to coincide.  The long-wavelength
spectral indices are +0.4 and +0.7, respectively.  The possible
interpretations of a moderately positive spectral index include jets
and winds \citep[e.g.][]{Reynolds86,Anglada98} and hypercompact HII
(HCHII) regions with steep radial density profiles \citep{Avalos06}.
In this case, 
we do not see any sign of spatial extension in the centimeter
wavelength emission for either of these sources -- they are both
unresolved with sizes $<0.3$\arcsec ($<400$~AU).  {Thermal radio
  jets from massive protostars have been seen to extend to scales as
  large as 0.1~pc \citep[e.g. IRAS~16547-4247]{Rodriguez05}. However,
  the small angular extent of the radio jet in the low-to-intermediate
  mass young stellar object HH~111~VLA~1 \citep{Gomez13,Rodriguez94}
  implies a size of only 116~AU, so the jet interpretation cannot be
  ruled out entirely on the basis of size alone. Nevertheless, the
  prior evidence that these are high-mass protostars \citep{Brogan09}
  leads us to favor the HCHII interpretation.}

To model HCHII region emission, we have
used the bremsstrahlung emission model V of \citet{Olnon75} in which
the electron density profile follows a power-law distribution ($n_e(r)
\propto r^{-2}$) transitioning to a sphere of constant density
embedded in the center, which avoids a non-physical singularity.  This
formulation yields a spectral index of +0.6 at frequencies below the
turnover point, which provides a good match to the SEDs of known
hypercompact HII regions \citep{Franco00}.  {To determine the
  optimal model, we used the ``lmfit'' Python
  package\footnote{http://lmfit.github.io/lmfit-py}, which performs
  non-linear least square minimization using the Levenberg-Marqardt
  algorithm.  Because the free-free emission and dust emission are
  comparable at frequencies just above the free-free turnover point,
  it is best to model both emission mechanisms simultaneously.  To
  model the dust emission, we use a single-temperature modified
  greybody function \citep{Rathborne10,Gordon95}.  With six flux
  density measurements, we attempted to model five free parameters,
  including three for the central sphere of the HCHII region: electron
  temperature ($T_e$), electron density ($n_{e0}$), and diameter
  ($d$); and two for the dust emission: the dust grain opacity index
  ($\beta$) and the opacity at the reference wavelength of 1.3~mm
  ($\tau_{\rm 1.3mm}$).  Because all of our measurements are on the
  Rayleigh-Jeans portion of the dust emission, we are unable to fit
  for the dust temperature; therefore, we fixed the dust temperature
  to the value determined for the gas (see \S~\ref{hotcore}), which is
  a valid assumption at the high densities of these sources.

  Unfortunately, simultaneous fits to all five parameters are not very
  constraining due to degeneracies between the parameters.  Instead,
  we defined a 16 by 16 point grid of electron density ($10^5$ to
  $10^8$ \percmc) and temperature (4000 to 14000~K), and fit the
  remaining three parameters at each grid point.  The grid points were
  uniformly distributed -- logarithmically for density and linearly
  for temperature.  For SMA~1b, 15\% of the results reproduced all the
  flux density measurements well and had rather similar $\chi^2$
  values.  For SMA~4, 23\% of the results were in this category.  The
  rest of the combinations of parameters were notably deviant at one
  or more data points.  In Figure~\ref{seds}, we plot the curve
  corresponding to the median parameter values and report the standard
  deviation of the five parameters across the good fits.}  For SMA~1b,
the dotted curve corresponds to $n_{e0} = 6 \times 10^5$ \percmc\/ and
a diameter of $d=44$~AU.  The expected full width at half maximum
(FWHM) of the total HCHII emission (i.e. including the portion outside
the central sphere) is then 54~AU.  The corresponding values for SMA~4
are $1.6 \times 10^7$ \percmc\/ and $d=7$~AU, yielding a FWHM of 9~AU.
Both of these sizes are consistent with being unresolved by our beam.
Therefore, one explanation for these objects is that they are
currently in the gravitationally-trapped phase of HCHII evolution
\citep{Keto07,Keto03}.  If so, then in the spherical accretion flow
model, the measured radius of an HCHII must be less than the
Bondi-Parker transonic radius, which in turn places a lower limit to
the central stellar mass \citep[see Equation 3 of][]{Keto07}.  Using a
sound speed of 11~\kms\/ for 10000~K gas, the corresponding lower
limits for the central stars in SMA~1b and SMA~4 are 4.4 and
1.4~\msun, respectively.  Of course, the presence of ionizing
radiation required to form the HCHII implies a central stellar mass of
$\sim$8~\msun\/ or more.  Evidently, the accretion flow is still
sufficiently strong to maintain the ionization radius significantly
inward of the point where pressure-driven expansion can begin, and
these stars can potentially continue to gain mass.



\subsubsection{SMA 1c}

In addition to a dust component, this source shows excess emission at
1.3~cm, the nature of which remains unclear. There is no significant
compact spectral line emission coincident with this source; the moment
images of HC$_3$N, OCS and $^{13}$CS give the impression that emission
avoids it and wraps around it (Figure~\ref{moments}).  More sensitive
observations at wavelengths longward of 2~cm are needed to understand
the nature of the centimeter continuum emission.  {To model the
  dust emission, we use the 7~mm and 1.3~mm flux densities and the
  assumed temperature range for the gas (20-50~K, \S~\ref{hotcore}).
  Because only two flux densities are available (as is the case for
  SMA13 and 15), we cannot fit for $\beta$ and $\tau_{\rm 1.3mm}$
  simultaneously. Instead, we first established a $\pm1\sigma$ range
  for $\beta$ by a Monte-Carlo treatment of the flux densities and
  their uncertainties.  We then derive the corresponding range of
  $\tau_{\rm 1.3mm}$ by solving the greybody model at each combination
  of the $T$ and $\beta$ extrema to find the value of $\tau_{\rm
    1.3mm}$ that matches the 1.3~mm flux density measurement. We
  report the minimum and maximum of the resulting values as the
  $\tau_{\rm 1.3mm}$ range in the text labels of Figure~\ref{seds}.}

\subsubsection{SMA 1d}

SMA~1d is perhaps the most intriguing source as it shows a consistent
spectral index from 1.3~cm to 0.87~mm.  {With measurements at four
wavelengths available, we used ``lmfit'' to fit the spectral index of
+2.25$\pm$0.07.}  This slope is only marginally steeper than an
optically-thick blackbody, and is similar to the values recently
reported in the low-mass class 0 protostar L1527 \citep{Tobin13} {and
in source VLA2 in the high mass star-forming core AFGL2591
\citep{vanderTak06}}.  If this emission arises from dust, then, as
summarized by \citet{Tobin13}, it could be explained by large dust
grains (compared to the wavelength), non-isothermal conditions, or
optically-thick structure, in which case it must be very compact.  A
similar spectral index is seen in the source NGC~6334~I-SMA4
\citep{Hunter13},
suggesting that it arises from a not uncommon phase of evolution of
massive star formation.  An alternative to dust is the core of an
optically thick jet \citep{Reynolds86}, in which the outer parts of
the jet are too faint to detect at the current sensitivity level.
Again, more sensitive and higher resolution observations are needed to
explore these possibilities.


\subsubsection{SMA 2, SMA 6, SMA 5, SMA 13 and SMA 15}

The rest of the continuum sources that are detected at 1.3~mm and
0.87~mm but not at longer wavelengths almost certainly arise from dust
emission.  If they were free-free emission, the allowed range of
spectral index (-0.1 for optically-thin emission to $+2$ for
optically-thick emission) would make them easily detectable in the VLA
images.  As described in \S~\ref{missingflux}, our 0.87~mm
measurements of SMA~13 and SMA~15 do not suffer from missing flux, and
their steep spectral indices from 1.3~mm to 0.87~mm are consistent
with thermal dust.  However, we must consider the chances of these
sources being extragalactic background objects dominated by dust
emission.  The South Pole Telescope (SPT-SZ) survey of 771 square
degrees in three frequency bands provides a good reference point for
comparison \citep{Mocanu13}.  From their cumulative distribution plot,
the number of SPT detections with dust-like spectral energy
distributions that are brighter than 25~mJy at 220~GHz is only
0.25-0.40 per square degree.  The probability of encountering such a
source in the field of view of Figure~\ref{fig1} is only $5 \times\/
10^{-6}$.  Thus, it is safe to conclude that these are not background
objects, but are members of the protocluster.  {In the case of
  SMA~2 and SMA~5, although the spectral indices are consistent with
  dust emission, we can only place lower limits on $\beta$: 1.0 and
  0.5, respectively (based on the upper limits at 1.3~cm and the lower
  limits at 0.87~mm).}

\subsubsection{VLA 2 and H$_2$O-C4}

Regarding the two new 6~cm point sources (VLA~2 and H$_2$O-C4) that
are not detected in millimeter continuum, the fact that one of them
(\water\/-C4) coincides with a \water\/ maser is compelling evidence
that it is tracing a protostellar object, {due to the strong
  association of molecular outflows with \water\/ masers
  \citep{Szymczak05,Codella04,Tofani95}}.  In fact, it is the fourth
brightest of the 11 \water\/ masers in the region, with an isotropic
luminosity of $1.6 \times 10^{-6}$ \lsun, which is characteristic of
massive young stellar objects (MYSOs) \citep[e.g.][]{Cyganowski13} and
the high end of intermediate-mass YSOs \citep[e.g.][]{Bae11}.  In the
\citet{Urquhart11} Red MSX Source (RMS) survey, which includes over
300 \water\/ maser measurements, the bolometric luminosities of MYSOs
with this maser luminosity range from $10^2-10^5$~\lsun, suggesting
that H$_2$O-C4 is a {\it massive} protostellar object.
%
%

\begin{figure}
\epsscale{1.0}
\footnotesize
\plotone{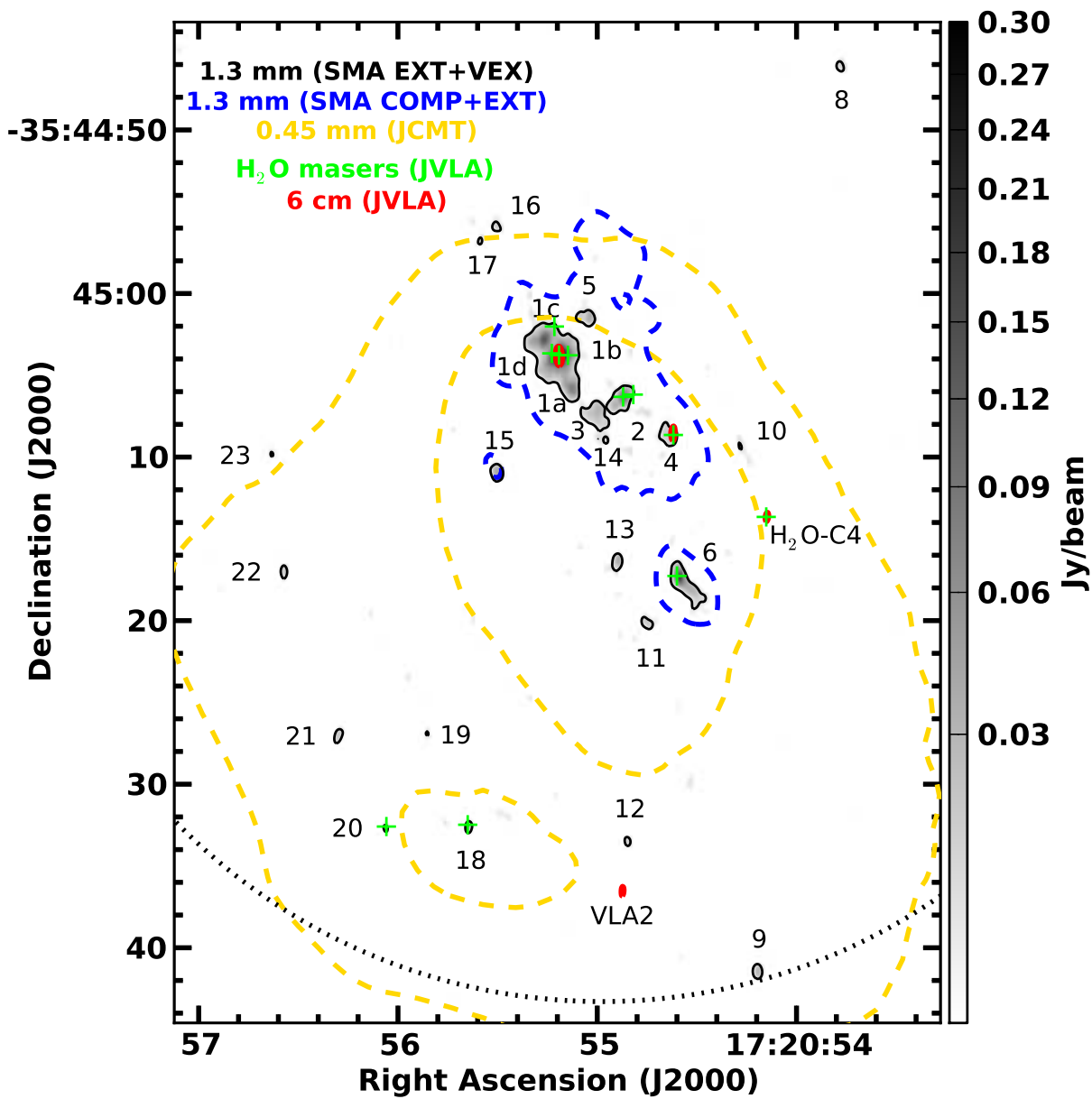} 
\caption{SMA 1.3~mm EXT+VEX continuum with $0.7\arcsec\times
  0.39\arcsec$ resolution in greyscale and a single black
  solid contour at $4.5\sigma\/ = 10$~\mjb. The dotted circle marks
  the 25\% level of the southern edge of the primary beam--although we
  searched for sources out to 20\% level, none were found beyond this
  radius. The labels correspond to the source names in
  Tables~\ref{contsources} and \ref{cmcontsources}. The image used in
  this figure was not corrected for the primary beam response, but all
  measurements, including Table~\ref{contsources}, were taken from the
  corrected image. The JVLA 6~cm continuum data is shown as a single
  red contour at 80~\mujb (4.5$\sigma$).  For ease of comparison with
  previous work and Fig.~1, the single dashed blue contour shows the
  40~\mjb\/ level of the COMP+EXT SMA 1.3~mm continuum and the dashed
  yellow contour shows the 60 and 100~\jb\/ levels of the JCMT 0.45~mm
  continuum. The green crosses mark the \water\ masers from
  \citet{Brogan09}.
\label{fig1}}
\end{figure}

\begin{figure}
\epsscale{1.0}
\plotone{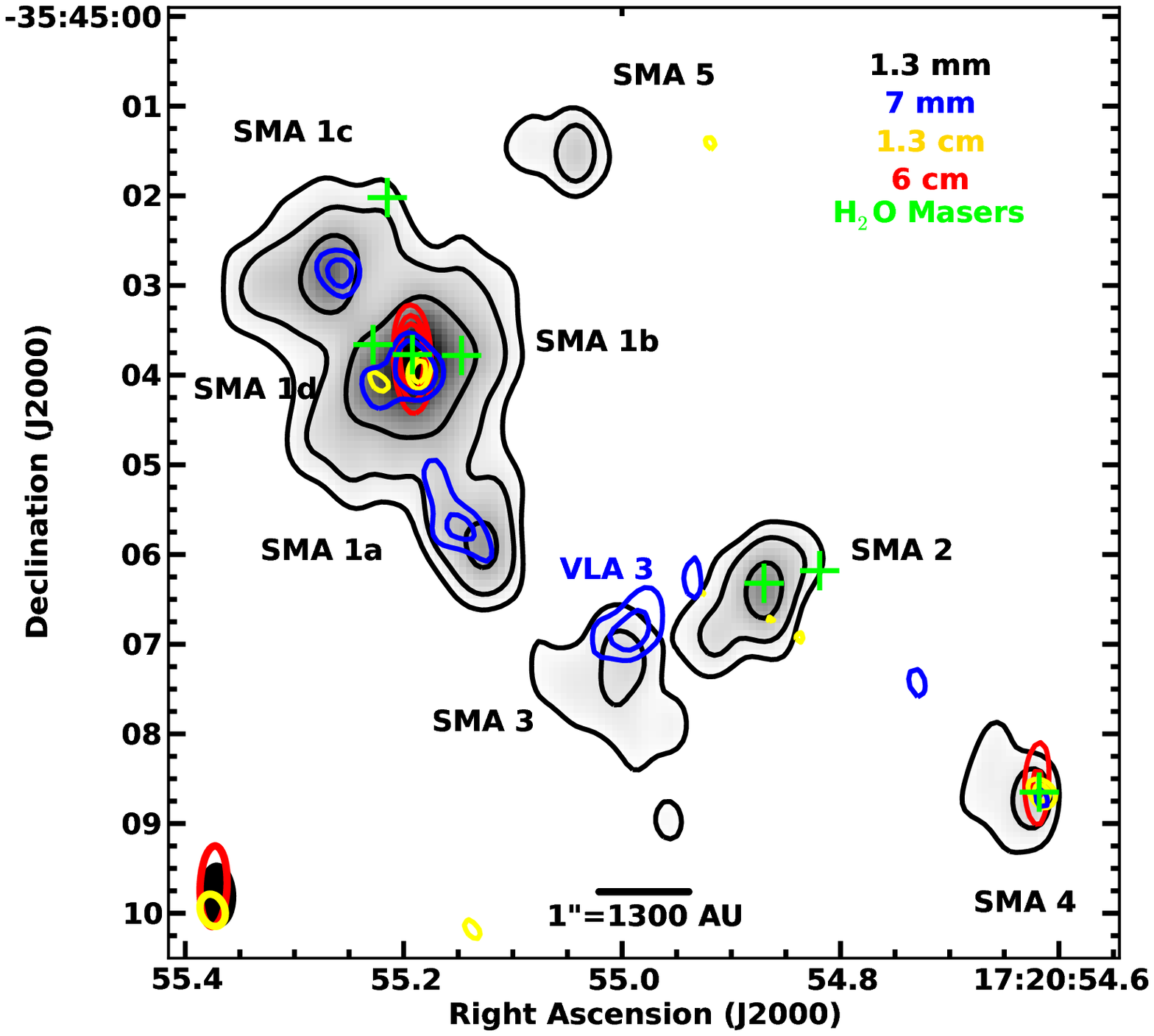} 
\caption{Zoom on the central portion of the \ngcin\/ protocluster
  including SMA1, SMA2, SMA3, and SMA4. The SMA 1.3~mm EXT+VEX continuum
  image is in greyscale and black contours (levels: 10, 24, and 66 \mjb).
  The red contours are the 6~cm VLA image (levels: 80, 144, and 200
  \mujb).  The blue contours show the 7~mm VLA continuum (levels:
  0.63, and 0.84 \mjb).  The yellow contours show the 1.3~cm VLA
  continuum (levels: 0.23, and 0.35 \mjb).  The green crosses mark the
  centroid positions of \water\/ masers from \citet{Brogan09}. The
  synthesized beams are shown in the lower left corner coded by contour
  color, the 7~mm beam is not shown as it is very similar to the SMA
  1.3~mm beam (black) (see Table~\ref{smaobs}).
\label{sma1zoom}}
\end{figure}

\begin{figure}
\epsscale{1.0}
\plotone{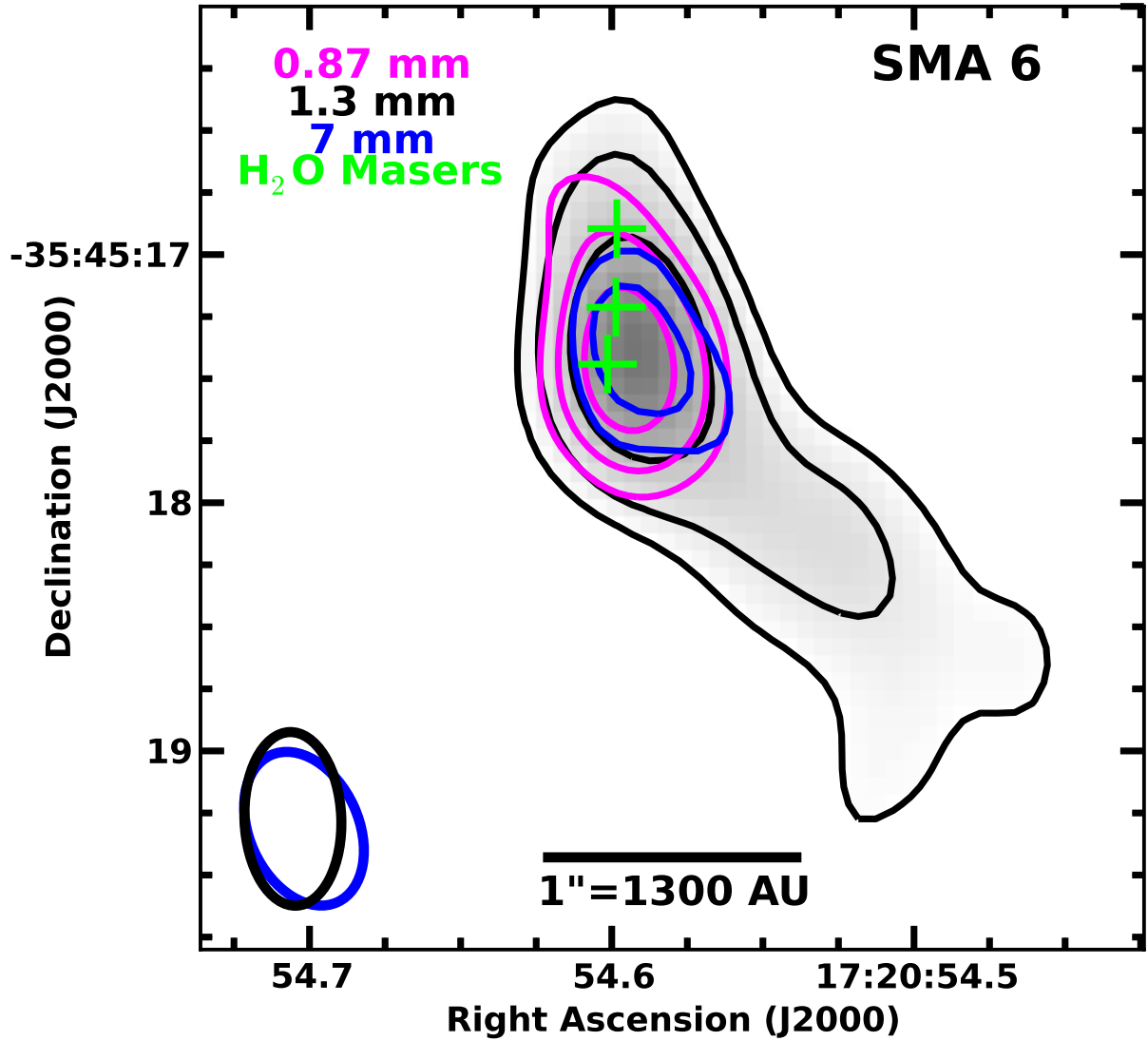} 
\caption{ Zoom on the SMA~6 region. The SMA 1.3~mm EXT+VEX continuum
  image is shown in greyscale and black contours (levels: 10
  {($4.5\sigma$)}, 24, 66 \mjb).  The magenta contours show the tapered
  and convolved (to match 1.3~mm) SMA VEX 0.87~mm continuum (levels:
  75 {($3\sigma$ at this point in the primary beam)}, 150, and 300
  \mjb).  The blue contours show the 7~mm VLA continuum (levels: 0.63
  {($3\sigma$)} and 95 \mjb).  The green crosses mark the centroid
  positions of the three groups of \water\/ masers from
  \citet{Brogan09}. The synthesized beams are shown in the lower left
  coded by contour color; the 1.3~mm and 0.87~mm resolutions are the
  same (black).
\label{sma6zoom}}
\end{figure}

\begin{figure}
\epsscale{1.0}
\plotone{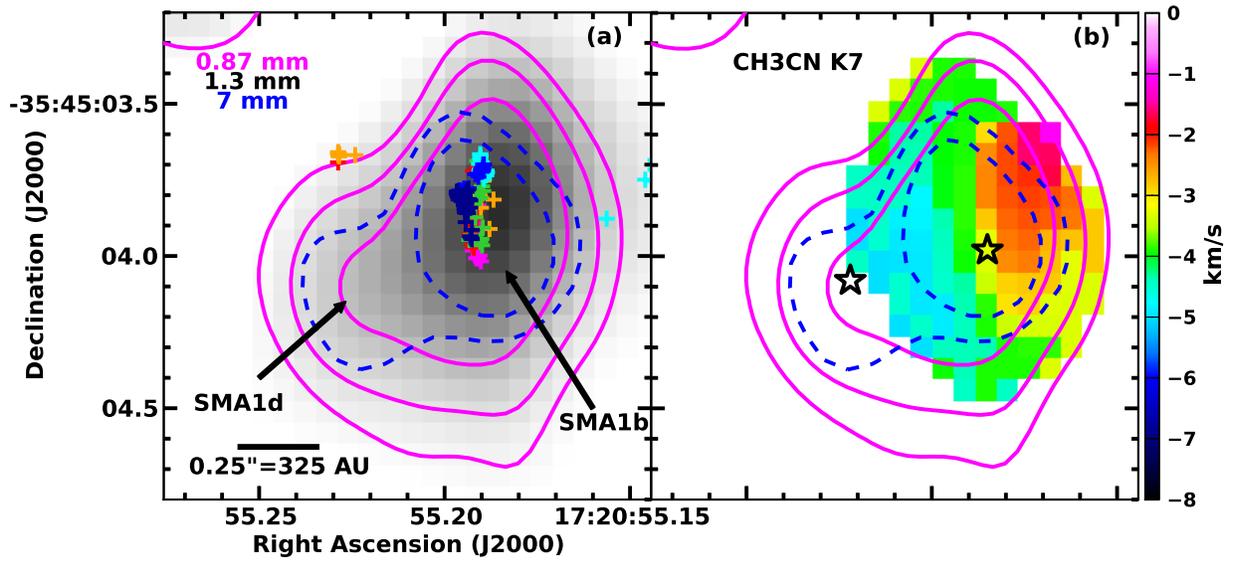} 
\caption{(a) The SMA 1.3~mm VEX-UV continuum image is in greyscale,
  and the colored crosses mark the positions and velocities of
  \water\/ masers from \citet{Brogan09}.  (b) The velocity field
  (moment 1) of the \methylcyanide\/ $J$=12--11, $K$=7 line is shown
  in colorscale, and the fitted positions of the 1.3~cm continuum
  emission (Table~\ref{cmpositions}) are marked by stars.  In both
  panels, the native resolution VEX 0.87~mm continuum image is shown
  by solid magenta contours (levels: 60, 120, 240~\mjb) and the 7~mm
  continuum emission is shown by dashed blue contours (levels: 0.63
  and 0.84~\mjb).
\label{sma1bd}}
\end{figure}

The other new source (VLA2) has no maser counterpart, so we must
assess the possibility that it is a background source.  Using the
formula of \citet{Anglada98} for the expected number of background
sources at 6~cm as a function of field size, we calculate a 4.4\%
chance of finding a background object of $>0.1$~mJy in the field of
view in Figure~\ref{fig1}.  Alternatively, using the source counts
from more recent EVLA 10~cm observations and extrapolating their flux
densities to 6~cm with the typical spectral index of $-0.7$
\citep{Condon12}, we find an expected source density of $4.5 \times
10^5$ sr$^{-1}$.  Multiplying by the solid angle of Figure~\ref{fig1},
we calculate a 6.3\% chance of finding a source in the range of
0.05-0.15~mJy.  Thus, the most likely scenario is that VLA2 resides in
the protocluster and may trace some form of young stellar object.
Given the lack of millimeter continuum and maser emission, the
centimeter emission could also arise from a low-mass, pre-main
sequence (PMS) star in which any associated circumstellar dust
emission is too faint to be detected.  Such objects have been seen,
for example in the Cepheus A East star-forming region
\citep{Hughes88,Garay96} at a distance of 700~pc \citep{Dzib11}.  The
variable sources (HW~3a, 8, 9) emit 0.15--3.0~mJy at 6~cm and show no
point source counterparts in SMA 0.87~mm images \citep{Brogan07}.  At
the distance of \ngcin, the 6~cm flux densities of these objects would
be 0.04--0.9~mJy, a range that includes both VLA2 and \water-C4.  It
is interesting to note that HW~8 and HW~9 are both X-ray sources with
HW~8 postulated to be a pre-main sequence star (due to its high median
photon energy), while the softer, brighter, and more variable HW~9 is
consistent with activity on a B-type star \citep{Schneider09}. In
\ngcin, a recent {\it Chandra} X-ray survey finds 7 sources in the
field of Figure~\ref{fig1}, but none of the detections correspond in
position to within 1.5\arcsec\/ of any of the millimeter or centimeter
sources, and their 1$\sigma$ X-ray position uncertainties are $\leq
0.3$\arcsec \citep{Feigelson09}.


\subsection{Protocluster structure and statistics}
\label{clusterstructure}

The unprecedented multiplicity of sources in this field allows us to
quantitatively analyze the structure of a young massive protocluster
for the first time. First, we define the membership of the
protocluster by requiring either the detection of compact millimeter
continuum emission or the positional coincidence of at least two star
formation indicators.  Starting with the source list of
Table~\ref{contsources}, we exclude SMA~7 as there is no compact
component detected.  Of the two new 6~cm sources, we can include only
the \water\/ maser source \water-C4.  VLA~2 could be protostellar, but
lacking a second positive indicator we exclude it.  Finally, although
we have presented strong evidence that SMA~1b and 1d are independent
sources, they may form a binary system.  Our sensitivity to binaries
is clearly limited by our angular resolution, so we choose to count
these two objects as a single source in order to avoid biasing the
following statistics by double counting some binaries but not others.
The final tally of sources is then $N_{\rm total}=25$.  The area of the
protocluster on the sky is limited by the 1.3~mm primary beam.  The
largest angular radius from the phase center of a detected source at
1.3~mm is $33''$, where $\sigma(r) \approx 5.4$~\mjb\/.  This implies
that our sample of the 1.3~mm compact sources in this region is
complete down to 25~mJy across a physical radius of 0.21~pc
corresponding to a projected area of 0.14~\pcsq\/ and a spherical
volume of 0.038~\pcc.  The presence of 25 sources within this volume
yields an average number density of 660~\ppcc. We note that only 20 of
the sources are above the 25~mJy completeness threshold; therefore,
this computed density is strictly a lower limit even if the underlying
population does not extend beyond the faintest detection.  In other
words, due to the radial decrease in sensitivity, the faintest sources
detected in the central portion of the field would have been missed in the
outer portions of the field.



Following the techniques applied to infrared and optical observations
of clusters of stars {and HII regions \citep[e.g.][]{Pleuss00,Schmeja06}},
we have constructed the minimal spanning tree (MST) formed from these
25 sources, as shown in Figure~\ref{mst}.  {The minimum spanning tree 
for a set of points is defined as the set of edges connecting them that 
possesses the smallest sum of edge lengths.  To compute it, we used a python
function based on the algorithm of \citet{Prim57}.}  From this
result, we compute the Q-parameter, which is defined as the ratio of
$\bar{m}$, the normalized mean edge length of the MST, to $\bar{s}$,
the correlation length, which is defined as the mean projected
separation between sources normalized by the cluster radius, $R_{\rm
  cluster}$ \citep{Cartwright04}.  $R_{\rm cluster}$ is defined as the
distance from the mean position of all the stars to the furthest star
from that point.  For this cluster, we obtain $R_{\rm cluster} =
31.95\arcsec$ and:
\begin{equation}
 Q = \frac{\bar{m}}{\bar{s}} = \frac{6.024\arcsec / \biggl(\frac{\sqrt{N_{\rm total}\pi R_{\rm 
  cluster}^2}}{(N_{\rm total}-1)}\biggr)} {19.93\arcsec/R_{\rm cluster}}  = \frac{0.510}{0.623} = 0.82
\end{equation}
Interestingly, this value (0.82) is close to the critical value that
separates clusters between the regime of multiscale (fractal)
substructure ($Q<0.8$) typified by the Taurus and Chamaeleon clusters,
and a centrally concentrated configuration typified by $\rho$~Oph and
IC348 \citep{Cartwright04}.  It is also close to the value of 0.84
obtained for stars in the Orion Trapezium Cluster \citep{Kumar07}.  In
the smoothed particle hydrodynamics simulation of the formation of a
1000~\msun\/ cluster \citep{Bonnell03}, the $Q$-parameter of the
resulting stars evolves steadily from initial values of $\sim$0.5 to
values greater than 1.1 \citep{Maschberger10}.  It passes the critical
0.8 value after about 1.8 free-fall times.  These results suggest that
our measured value for $Q$ for this protocluster is not a fixed
property but may instead provide a measure of its age.  However, we
caution that our census of the cluster may be limited by the primary
beam of the observations, i.e. the cluster may extend beyond this
region in one or more directions.  It is also limited in terms of
sensitivity depth. Future more sensitive observations with a wider
field of view are needed to improve the robustness of this result.  In
any case, it is likely that this protocluster exists in a stage when
its ultimate structure is still to be determined.

\subsection{Dynamical mass and relaxation time}

\label{dynamicalmass}

Using the virial theorem, the dynamical mass of a stellar cluster can
be estimated from the mean square velocity of its members (relative to
their mean velocity) as originally performed on the globular cluster
M92 by \citet{Wilson54}. For the \ngcin\/ protocluster, we compute the
{unbiased sample} variance of the source-to-source LSR velocities: 
$\langle v_{\rm
  1D}^2\rangle = (1/(N_{\rm src}-1))\Sigma (v_{\rm src} - \bar{v})^2$, by
using the fitted velocities ($v_{\rm src}$) from the single-component
\methylcyanide\/ spectral models of the $N_{\rm src} = 6$ sources in
Table~\ref{temperatures} and their mean ($\bar{v}$).  {As seen in
  Figure~\ref{mst}, three of these sources lie in the central strip, two
  in the first pair of flanking strips, and one in the second pair of
  flanking strips.  Thus, although small in number,} these six sources
provide a {fair} spatial sampling across the cluster.  The
\water\/ maser emission provides a potential kinematic measurement for
two additional sources: SMA~20 and H$_2$O-C4.  However, \water\/
masers typically span a broad range in velocity, and the centroid
can vary widely from the LSR of the thermal gas \citep{Urquhart11}, as
is the case for SMA~18, which is detected in both \methylcyanide\/ and
\water.  Therefore we have chosen not to include these two sources in
this analysis.
{The resulting value and uncertainty of the unbiased sample variance
  are $\langle v_{\rm 1D}^2\rangle = 2.05 \pm 1.29$~\kmssq, where the
  uncertainty in the variance is the variance times $\sqrt{2/(N_{\rm
  src}-1)}$ \citep[e.g.][]{Casella02}.} Assuming a random orientation
  of space velocities, the mean square three dimensional velocity is
  then $\langle v_{\rm 3D}^2\rangle = 3\langle v_{\rm 1D}^2\rangle =
  6.15 \pm 3.87$~\kmssq.  The proper effective distance to use in the
  virial theorem, $\bar{r}$, can be determined by numerical
  integration of the observed strip counts \citep{Schwarzschild55}.
  In this case, the strip counts are the number of stars in each
  horizontal bin shown in Figure~\ref{mst}.  We chose a bin width of
  10\arcsec\/ as it provides seven bins, which is a good compromise
  between having too few bins and having too few sources per bin.
  Using equation 1 of \citet{Schwarzschild55}, we obtain $\bar{r} =
  44.3\arcsec$, or 0.28~pc, yielding a dynamical mass estimate of $410
  \pm 260$~\msun.  By comparison, the gas mass of \ngcin\/ plus SM2
  based on single-dish submillimeter continuum observations is
  $280$~\msun, assuming a uniform dust temperature of 30~K
  \citep{Sandell00}, after scaling to the new distance of 1.3~kpc.
  Because the \citet{Sandell00} mass is based on Gaussian fits to
  these two sources, it is a lower limit to the total gas mass as it
  does not include the mass of the more extended material between them
  (see Figs.~\ref{fig0} \& \ref{fig1}).  Furthermore, the total gas
  mass is a lower limit to the total cluster mass because it does not
  include the mass of condensed (proto)stars.  Considering these
  effects, the agreement between these two measurements is remarkable,
  and lends credence to the interpretation of the hot core velocity
  dispersion as tracing the dynamics of the protocluster.  We note
  that the velocity dispersion $\sigma_v = \sqrt{\langle
  v_{1D}^2\rangle} = {1.43}$~\kms\/ is in good agreement with the
  internal radial velocity dispersion of the Scorpius OB2 cluster
  (1.0-1.5~\kms) as measured by kinematic modelling of Hipparcos data
  \citep{deBruijne99}.

We next consider the relaxation time of the protocluster, which is
based on the crossing time:
\begin{equation}
t_{\rm cross} = R_{\rm cluster}/v_{\rm 3D} = 87000 \mbox{ yr}.
\end{equation}  
Because the number of known protostars is still fairly small, 
the relaxation time \citep[see e.g.][]{Binney87} is essentially 
the same:
\begin{equation}
t_{\rm relax} = t_{\rm crossing} \frac{N_{\rm total}}{8 \ln(N_{\rm total})} = 84000 \mbox{ yr}.
\end{equation}  
Given the likelihood that the protostellar population continues below
our sensitivity limit, this value should be considered an upper limit
to the relaxation time.  The low value of the $Q$-parameter suggests
that the protocluster is dynamically young in the context of
evolutionary simulations \citep{Parker14}.  However, the good
agreement between the dynamical mass and the gas mass suggests that
the protocluster has already persisted for one or more crossing times,
which theory suggests is sufficient for dynamical mass segregation to
occur \citep{Bonnell98}.  Although the number of detected objects is
too small to apply a statistical estimate of mass segregation
\citep{Allison09}, the probable most massive source in the
protocluster (SMA~1b) is by no measure located near the center of the
distribution of members (see Figure~\ref{fig1}).  Therefore, we do not
see any clear evidence for mass segregation.


\subsection{Individual dust masses: circumstellar disks?}
\label{dustcalc}
Our single-component LTE model fits (Table~\ref{temperatures}) have
provided gas temperature and column density measurements for six of
the sources.  Given the high volume densities implied by these fits,
we make the usual assumption that the dust temperature ($T_{dust}$) is
well coupled to the gas temperature through collisions.  To obtain gas
mass estimates from the 1.3~mm dust emission, we follow the procedure
of \citet{Brogan09} using the dust mass opacity coefficient
$\kappa_{\rm 1.3mm} = 1$~cm$^{2}$~g$^{-1}$ appropriate for grains with
ice mantles in regions of high gas density (10$^8$~\percmc)
\citep{Ossenkopf94}, {and a gas to dust mass ratio of 100}.  In
the sources that do not show \methylcyanide\/ emission, we have no
temperature measurement.  In these cases, we have assumed a range of
temperatures that are fairly low (20--50~K), and compute the range of
masses implied by the measured flux density.  This temperature range
encompasses both the typical temperature of gas surrounding young
protostars prior to the hot core phase, as well as the expected mean
temperature of gas in circumstellar disks around stars of type A3 --
B5 \citep{Natta00}.  The masses were corrected for the dust opacity by
comparing the observed brightness temperature to the assumed dust
temperature ($\tau_{dust}=-\ln(1-T_b/T_{dust})$) and the correction
factor is: $\tau_{dust}/(1-\exp(-\tau_{dust}))$. The resulting mass
estimates for all sources are shown in Table~\ref{dustmass}, and for a
subset of sources in Figure~\ref{seds}.

With the exception of SMA~1a,b,c,d, SMA~3, and SMA~9, the mass
estimates are in the range of 0.2-2~\msun.  These masses cover a range
similar to the masses of dense cores in the Perseus molecular cloud
\citep{Kirk06}; however, the smallest Perseus objects have a typical
diameter $\approx$5000~AU (20\arcsec\/ at 250~pc) in submillimeter
continuum images \citep{Kirk07}.  Such objects would be nearly
4\arcsec\/ in diameter at the distance of \ngcin.  Therefore, the
emission we are detecting must arise from structures that are an order
of magnitude (or more) smaller, which includes circumstellar disks.
The mass values are not as high as massive disk candidates
\citep[e.g. the 0.8-8\msun\/ of IRAS~20126+4104,][]{Cesaroni05}, but
they are an order of magnitude higher than the typical masses of
circumstellar disks around low-mass stars, for example
0.005-0.14~\msun\/ in the Ophiuchus sample of \citet{Andrews09}.
However, they are not too massive to be disks around intermediate mass
stars.  There are a number of examples of Herbig A stars with disk masses
of $\sim$0.2\msun, such as Mac~CH12 \citep{Mannings00}.  Its 1.3~mm
flux density of 44~mJy \citep{Osterloh95} would be 19~mJy at the
distance of \ngcin.  A more massive example is SMA1 in G5.89-0.39,
which appears to be an intermediate mass protostar with
$\sim$1~\msun\/ of circumstellar material \citep{Hunter08}.
Thus, {along with the lower mass population of disks around
  high-mass protostars \citep[such as the 0.8~\msun\/ of AFGL2591~VLA3;][]{vanderTak06},} we may also be detecting the most massive tip
of the population of disks around intermediate-mass protostars in this
protocluster.

Finally, in principle one can use the H$_2$ gas column density
estimated from the dust column density to calculate the
\methylcyanide\/ abundance, subject to the assumption that both
tracers arise from the same volume of gas. In the case of SMA~1b, the
fitted size of the dust emission (0.87\arcsec\/ at 1.3~mm and
0.47\arcsec\/ at 0.87~mm) is significantly larger than the modeled
size of the \methylcyanide\/ emission region (0.38\arcsec), so this
assumption is not valid.  Nevertheless, proceeding with this
uncertainty, the gas column density toward SMA~1b inferred from the
dust emission is $1 \times 10^{25}$~\percms, yielding an abundance
estimate of $4 \times 10^{-8}$.  Similarly, for the next two brightest
line sources (SMA~2 and SMA~4), we obtain abundances of $5 \times
10^{-9}$ and $6 \times 10^{-8}$.  Given all the uncertainties, these
values are in reasonable agreement with the assumed value of $1 \times
10^{-8}$ (see \S~\ref{hotcore}). Higher resolution observations are
needed to obtain more accurate estimates of the molecular abundances
via this method.





\subsection{A candidate massive rotating disk in SMA~1b}

The position angle of the velocity gradient in the molecular gas in
SMA~1b ($-52\pm$5\arcdeg) is equal to the position angle of the
two-dimensional Gaussian model fit to the dust emission
(Figure~\ref{sma1bd}): $-52\pm$9\arcdeg at 1.3~mm
(Table~\ref{contsources}) and $-51\pm$8\arcdeg at 0.87~mm
(Table~\ref{contsourcesBand7}).  Furthermore, this position angle is
91\arcdeg\/ different from the position angle of $+39$\arcdeg\/
determined for the SiO~5-4 outflow \citep[][also see
  Figure~\ref{methanolmasers}]{Brogan09}.  This perpendicularity is
heavily suggestive of a disk-outflow system.  {In a few of the
  lower temperature transitions, including \hcccn, OCS, and $^{13}$CS,
  emission extends beyond SMA~1b by up to an arcsecond along the
  outflow direction on one or both sides. This behavior is not
  expected from a compact disk.  However, at these locations, the
  kinematic structure changes from a strong gradient to a narrower
  range of velocities, suggesting that these transitions are 
  tracing an interaction of the outflow with ambient material near the
  disk.}

\begin{figure}  
\epsscale{1.0} 
\plotone{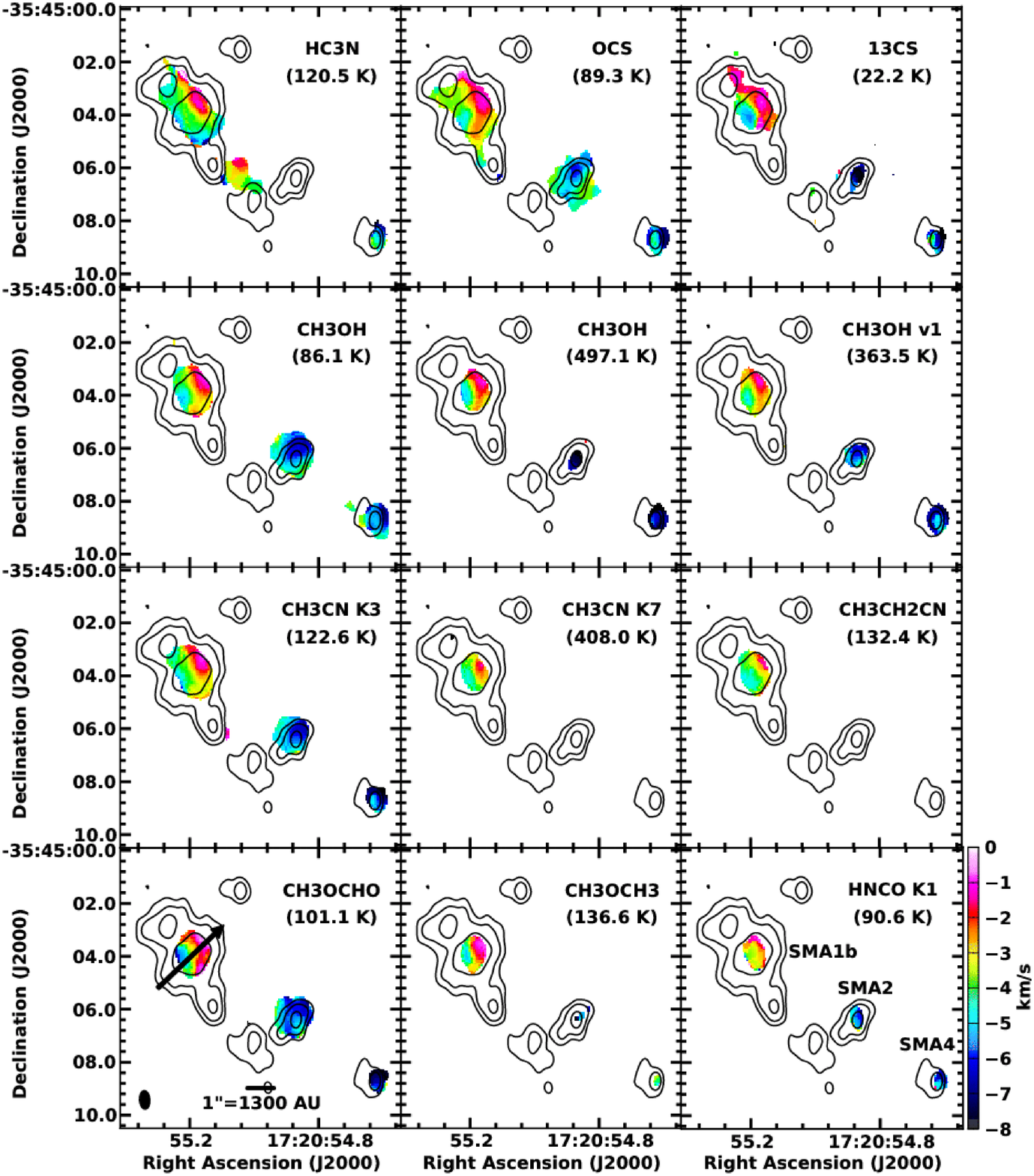} 
\caption{Images of the first moment of the 12 spectral lines listed in
  Table~\ref{linelist} from the SMA VEX 1.3~mm data.  The field of
  view is the same as Figure~\ref{sma1zoom}. The excitation energy (in
  Kelvin) of the lower level of each transition is indicated.  The
  black contours show the 1.3~mm EXT+VEX continuum emission (contour
  levels: 10, 24, and 66 \mjb\/). The arrow in the lower left panel
  indicates the cut used to generate the pv diagrams in
  Figure~\ref{pv}.  The resolution of the EXT+VEX continuum and VEX
  line data is similar (see Table~1); the beamsize is shown in the
  lower left corner of the lower left panel.
  \label{moments}
}
\end{figure}

\begin{figure} 
\epsscale{1.1} 
\plotone{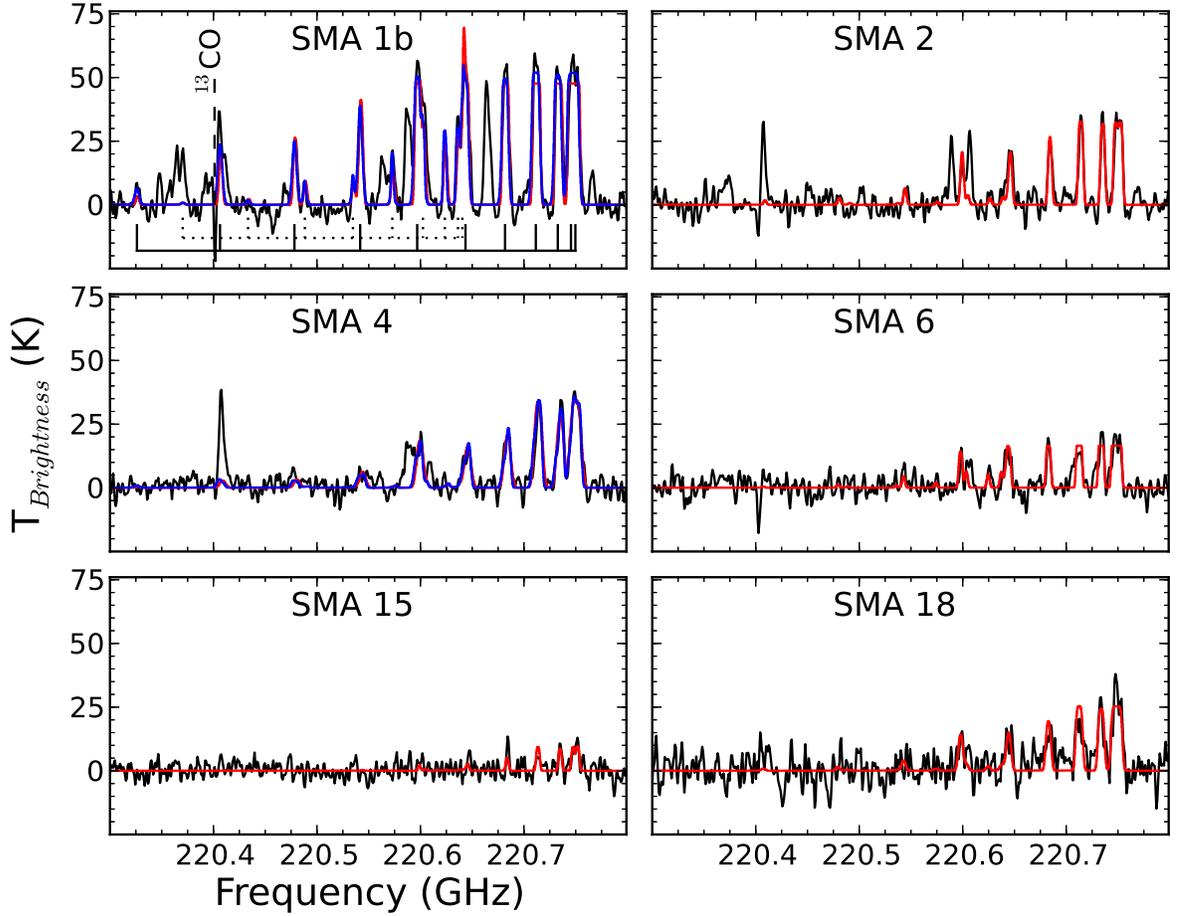} 
\caption{In each panel, the black spectrum is the observed spectrum
  toward the indicated source in units of brightness temperature
  (corrected for primary beam response), and the red spectrum is the
  best fit single-component CASSIS LTE model of \methylcyanide\ and
  \ttmethylcyanide\/ $J$= 12--11.  The blue spectrum shown for SMA~1b
  and SMA~4 is a two component model.  The corresponding model
  parameters are given in Table~\ref{temperatures}.  In the SMA~1b
  panel, the vertical lines below the spectrum mark the frequencies of
  the $K$ components of \methylcyanide\/ (solid lines) and
  \ttmethylcyanide\/ (dotted lines). The rest frequency of $^{13}$CO
  is also marked by the vertical dashed line.  The flat tops on many
  of the low $K$ lines are indicative of high optical depths.
  \label{linefits}
}
\end{figure}

\begin{figure} 
\epsscale{1.0} 
\plotone{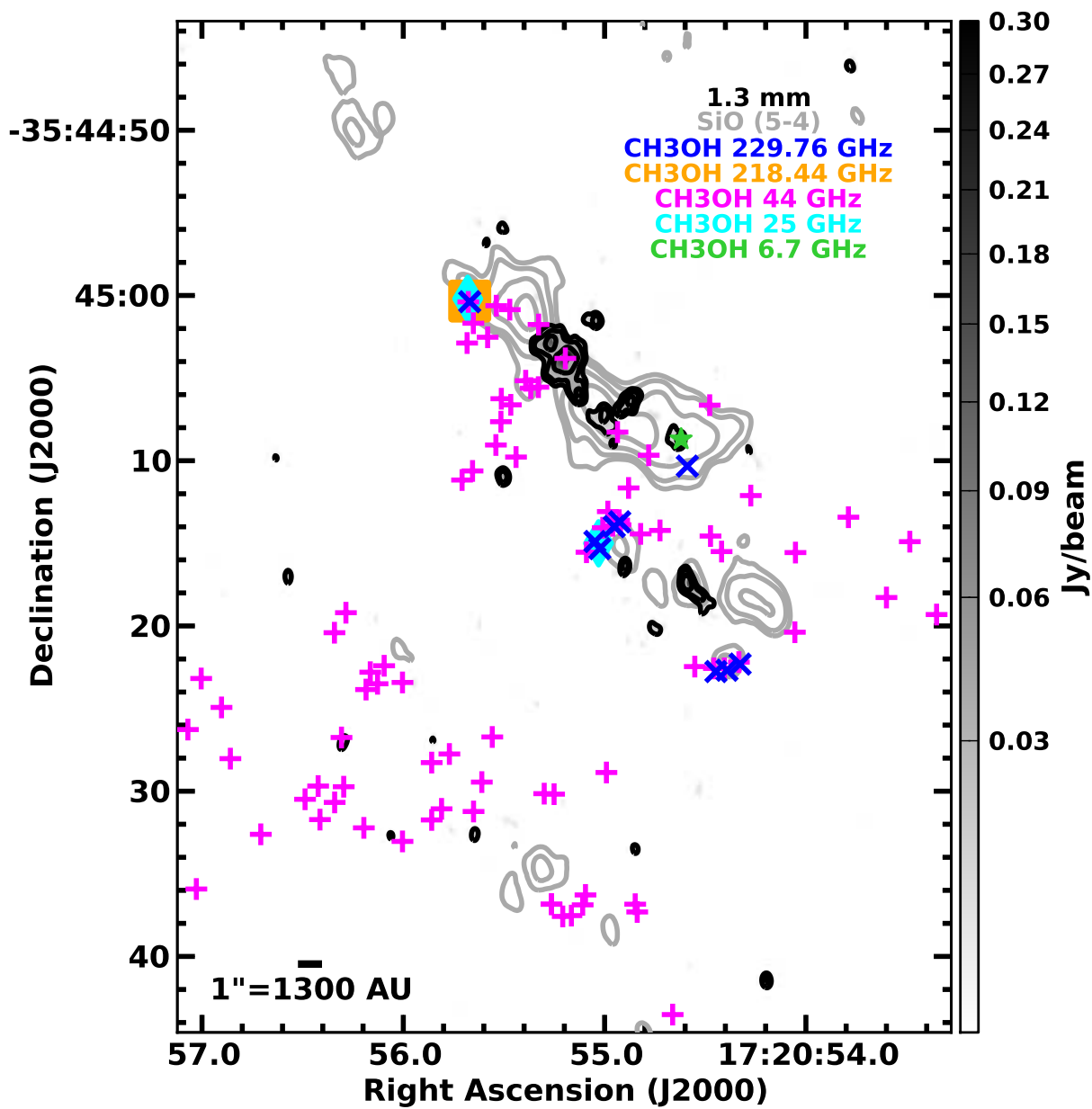} 
\caption{Image of the EXT+VEX 1.3~mm continuum in greyscale and black
  contours, (same three contours as in Figure~\ref{sma1zoom}).  The
  field of view is identical to Figure~\ref{fig1}.  The integrated
  intensity of the SiO $J$=5--4 emission from the COMP+EXT SMA data
  presented in \citet{Brogan09} is shown in gray contours, indicating
  strong outflow emission.  The positions of the various \methanol\/
  masers are marked as follows: $\times$: 229.76~GHz; square:
  218.44~GHz; $+$: 44~GHz; diamond: 24.9~GHz \citep{Beuther05};
  5-pointed star: the intensity-weighted centroid of the six 6.7~GHz
  masers (17:20:54.619, -35:45:08.66) computed from the table in
  \citet{Walsh98}.  The positions and velocities of the newly
  discovered Class I 218.44 and 229.76~GHz \methanol\/ masers are
  tabulated in Table~\ref{masertable}.
  \label{methanolmasers}
}
\end{figure}

\begin{figure}   
\epsscale{0.7} 
\footnotesize
\plotone{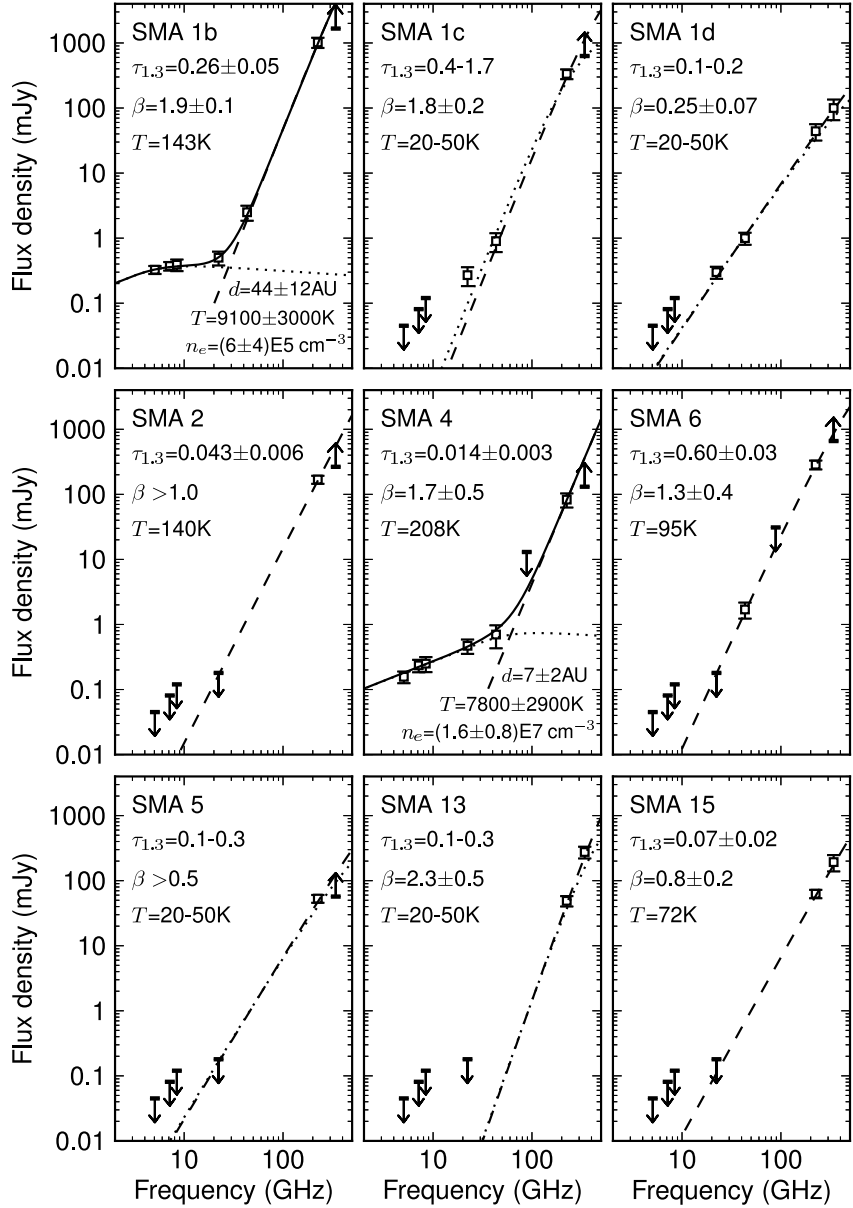} 
\caption{\small The spectral energy distributions of the nine
  continuum sources detected at more than one wavelength. The 6~cm,
  1.3~mm and 0.87~mm measurements are from this paper, while the rest
  can be found in \citet{Brogan09}.  The flux density error bars
  represent the formal error of the two-dimensional Gaussian model fit
  plus a fraction of the total flux density (to account for the
  typical calibration uncertainty, see \S~2). Non-detections are shown
  as 3~$\sigma$ upper limits. The 3~mm flux density measurements for
  SMA~4 and SMA~6 \citep{Brogan09} are regarded as upper limits for
  the size scales considered here.  The lower limits drawn for the
  detections at 0.87~mm are due to the relative lack of short spacings
  compared to the 1.3~mm measurements.  In SMA~1b and SMA~4, the
  dotted curve is a free-free emission model \citep[$n_e\propto
    r^{-2}$:][]{Olnon75} whose parameters are listed in the lower
  right, the dashed curve is the dust model, and the solid curve is the
  sum.  The parameters of the dust emission model are listed in the 
  upper left of each panel.  For SMA~1c, 1d, 5, and 13, the dashed 
  and dotted curves are $T$=50~K and $T$=20~K dust, respectively.
\label{seds}}
\end{figure}

\begin{figure} 
\epsscale{0.8} 
\plotone{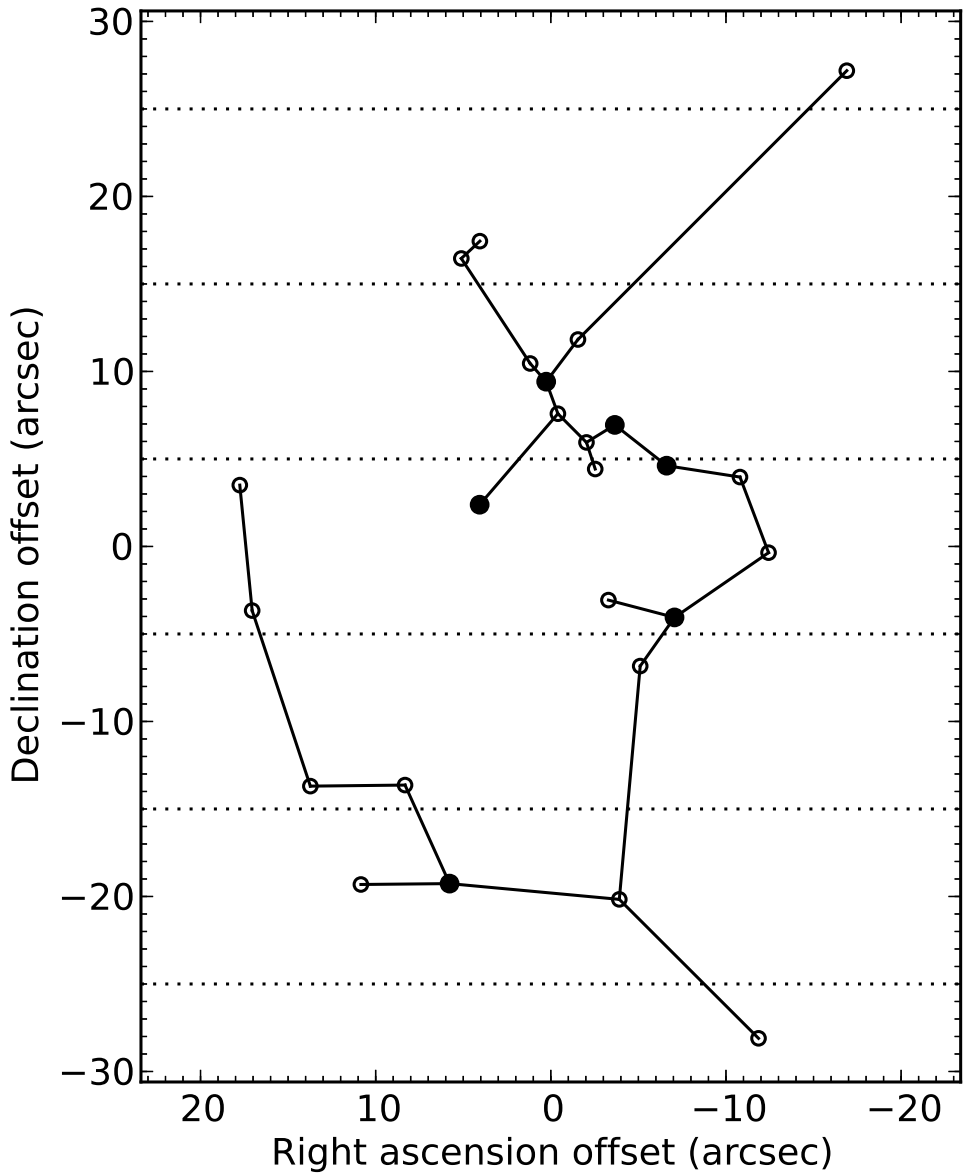} 
\caption{The minimum spanning tree of the protocluster as defined by
  the 24 detected sources in Table~\ref{contsources} plus the water
  maser/6~cm source \water-C4 from Table~\ref{cmcontsources}.  The
  origin is the mean position of all sources.  The filled points are
  the six sources for which we were able to measure the LSR velocity
  (Table~\ref{temperatures}).  The horizontal dashed lines demarcate
  the seven 10\arcsec-wide strips in which the protostars were counted
  (from north to south: 1, 2, 6, 9, 3, 3, 1) in order to compute the
  effective distance to be used in the calculation of the dynamical
  mass in \S~\ref{dynamicalmass} by the method of \citet{Schwarzschild55}.
  \label{mst}}
\end{figure}

\begin{figure} 
\epsscale{0.9} 
\plotone{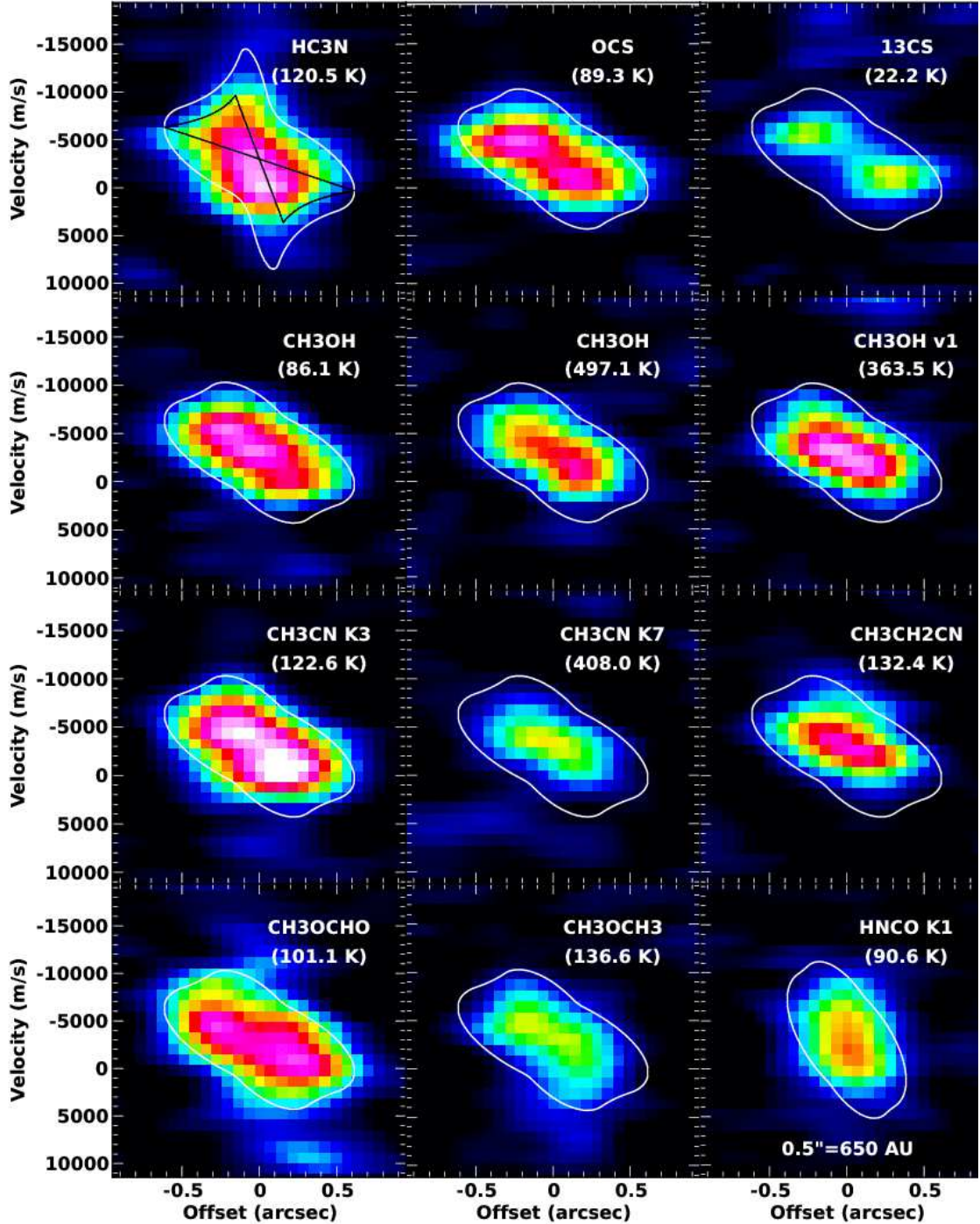} 
\caption{\small Position-velocity diagrams taken along the position angle of
  the disk ($-52\arcdeg$) for the same molecular transitions shown in
  Figure~\ref{moments}.  The intensity map is the same in all panels
  {(0-0.6~\jb)}.  In all panels, the white contour encompasses the
  region where emission is expected from an edge-on disk 
  at an LSR velocity of $-3.0$~\kms\/ with a
  central enclosed mass of 10~\msun\/ undergoing Keplerian rotation and
  free-fall.  In the HC$_3$N panel,
  the parameters of the model are $r_{\rm inner}$=200~AU and $r_{\rm
    outer}$=800~AU, {and the black contour shows the model without
    free-fall}.  In the HNCO panel, the parameters are $r_{\rm
    inner}$=400~AU and $r_{\rm outer}$=500~AU, and in the rest of the
  panels, the parameters are $r_{\rm inner}$=500~AU and $r_{\rm
    outer}$=800~AU.
\label{pv}}
\end{figure}

To further explore the velocity gradient toward SMA~1b, we constructed
position-velocity (pv) images (Figure~\ref{pv}) for the same 12
transitions whose moment maps were shown in Figure~\ref{moments} {(the
position and orientation of the slice is shown on the CH$_3$OCHO panel
of Figure~\ref{moments}).  In the context of a disk, three main
factors will govern the appearance of the pv image of a molecular
line: (1) the overall physical structure of the circum-protostellar
material, in particular the outer radius, inner radius (if there is a
central cavity), and the thickness; (2) the radial gradient in the
temperature, density, and molecular abundance; and (3) the observer's
viewing angle.  Evidence for chemical segregation in the context of a
hot core was recently reported for the inner 3000~AU of the
AFGL2591~VLA3, in which different species appear to trace different
radii with respect to the continuum peak \citep{Jimenez12}.  One
interpretation of this object is that it contains a massive disk
\citep{Hutawarakorn05,Trinidad03}.  In the case of \ngcin\/~SMA~1b},
we observe the overall shapes of the structures in the pv images to be
generally consistent with one another, but the differences are worth
noting.  {In general,} the high temperature transitions are peaked
toward the central source, {including the two highest temperature
\methanol\/ lines, \methylcyanide\/ K=7, and CH$_3$CH$_2$CN. In
contrast, most of} the lower temperature transitions show a double
peak with a local minimum toward the central source, {including
$^{13}$CS, OCS, CH$_3$OCHO and \methylcyanide\/ K=3}.  This pattern is
consistent with the combination of a radial temperature gradient in
which the hotter gas lies closer to the central heating source, {and a
viewing angle that is edge-on (or at least moderately so). The two
species that deviate from this picture are HNCO, which shows a
distinctive compact morphology, and \hcccn, which extends to higher
velocities.}

{Regarding HNCO, \citet{Tideswell10} recently modelled its
  abundance in hot cores versus time using a wide range of assumed
  chemical reactions (both grain and gas-phase), along with different
  initial cloud collapse temperatures and post-collapse (hot core)
  temperatures. They find the best agreement with observed abundances
  from models that include both grain and gas-phase chemistry, cloud
  collapse temperatures of 10~K, and post-collapse temperatures
  $>50$~K. Furthermore, the late peak of HNCO abundance with time ($2
  \times\/ 10^5$ yr) suggests that HNCO is not itself a ``first
  generation'' species liberated from dust grains. Instead, it is the
  ejection of several larger HNCO ``daughter'' molecules formed on the
  grains (such as HNCOHO, HNCOCHO, HNCONH, and HNCOOH) and their
  subsequent destruction that form HNCO in the gas phase. The compact
  size of the observed HNCO emission relative to other species
  suggests that this formation scenario is efficiently proceeding only
  in the inner disk, perhaps due to the higher temperatures there.}

Regarding \hcccn, not only does its emission extend to higher
velocities, but these velocities arise exclusively from locations
close to the center.  This pattern suggests faster rotation toward the
inner radii. {However, it is somewhat surprising that HC$_3$N is
  seen closer to the central source than other molecules given that it
  has a relatively high photodissociation rate \citep{Martin12} and
  that ionizing radiation should be stronger at smaller radii.  On the
  other hand, if there are locations with sufficient shielding, such
  as in the mid-plane of a high column density disk, the formation of
  HC$_3$N can proceed quickly as a product of C$_2$H$_2$ and CN
  \citep[e.g.][]{Meier05,Fukuzawa97}, and CN has a low
  photodissociation rate \citep{Martin12}. It is interesting to note
  that the \hcccn\/ abundance observed toward the continuum peak of
  AFGL2591~VLA3 by \citet{Jimenez12} exceeds their gas-grain chemical
  evolution model predictions by more than two orders of magnitude.
  Indeed, these authors emphasize that molecular abundances are very
  sensitive to extinction, and hence the detailed geometry of the
  system.  }


As seen in Figure~\ref{pv}, the maximum extent of the emission toward
SMA~1b is $\pm0.6$\arcsec\/ ($\pm$800~AU) and the observed range of
gas velocities at that radius is approximately $\pm3.5$~\kms\/ with
respect to the LSR.  If this emission is interpreted as arising from a
circular rotating disk, then these two quantities alone can be used to
make a simple dynamical estimate of the central enclosed mass using
Kepler's law for circular rotation: $M_{\rm enclosed}=R(V/\sin i)^2/G
= 11/\sin^2i$ \msun\/, where $i$ is the angle of inclination of the
rotation axis to the line of sight. However, rotation is not the only
possible motion, as evidence for infall in this source has already
been demonstrated by the trend of the increasingly redshifted
absorption features with increasing line excitation seen in three
transitions of CN and \formaldehyde\/ \citep{Brogan09}.  Therefore, to
build a more illustrative physical model, we follow the approach of
Equation~1 of \citet{Cesaroni11}, where we interpret the pv images as
arising from a rotating Keplerian disk undergoing free-fall and having
a specific inner and outer radius.  Using this model, the white
contour in Figure~\ref{pv} encompasses the entire region where
emission can arise (ignoring any effects of line opacity and dust
opacity) from a disk with an enclosed mass of $M_{\rm
  enclosed}=10/\sin^2i$~\msun, and the specified inner and outer radii
for each transition. {(For comparison, the model without free-fall
  is included as the black contour in the first panel of
  Figure~\ref{pv}.)}  The inner radius sets the total extent of the
white contour along the velocity axis, while the outer radius sets its
total extent along the position axis.  Most of the molecular
transitions are consistent with a model having an inner radius of
500~AU, and outer radius of 800~AU.  The HC$_3$N line is more
consistent with a smaller inner radius of $\sim$200~AU, while the HNCO
line is more consistent with originating from a narrow range of inner
radii (400-500 AU).

The outer radius of 800~AU seen in most species is comparable to the
semi-major axis of the deconvolved Gaussian model of the continuum
from SMA~1b, which is 660$\pm$33~AU at 1.3~mm
(Table~\ref{contsources}) and 403$\pm$20~AU at 0.87~mm
(Table~\ref{contsourcesBand7}).  For example, an outer radius of
800~AU for the molecular gas would correspond to the 7$\pm$2 percent
point of the 0.87~mm model dust source.  Thus, the extent of the dust
and gas emission yield similar values for the size of the structure.
Assuming circular symmetry, the observed ellipticity of the 0.87~mm
deconvolved model can then be used to set a lower limit for $i$ of
55\arcdeg, i.e. nearly edge-on.  The corresponding upper limit for the
enclosed mass is then 30~\msun.  We note that for a disk viewed
edge-on, the line emission from the innermost radii will be obscured
by the high optical depth of the foreground material in the disk, both
from the gas and (at some point) the dust grains.  To demonstrate, the
\methylcyanide\/ optical depths at line center for the $K$=7 and 8
components, derived from their peak brightness temperatures in the
single-temperature CASSIS model, are 0.7 and 0.4, respectively.
Although the presence of a Keplerian disk is tantalizing, future
higher resolution observations are necessary to properly model the
geometry and inclination angle and to confirm the expected kinematic
structure in more detail.  Radiative transfer modelling will no doubt
be profitable in delineating and interpreting the chemical structure
as it has been for low-mass protostellar disks \citep{Qi08}.


\section{Conclusions}

Using new high-resolution SMA and VLA images from 6~cm to 0.87~mm, we
have found further multiplicity in \ngcin, allowing us to perform the
first structural analysis of a massive protocluster using techniques
developed for optical/infrared studies of star clusters.  We also
demonstrate the first use of the thermal gas velocities from an
ensemble of hot cores to probe the dynamical properties of a
protocluster.  Our results are summarized as follows:

\begin{itemize}

\item We have identified 16 new compact 1.3~mm continuum sources and
  two new 6~cm sources. Three of the newly discovered sources are
  associated with \water\/ masers. Combined with the previously-known
  1.3~mm sources from \citet{Brogan09}, the total number of compact
  centimeter or millimeter sources is 28.  Limiting our analysis to
  likely protostars, i.e. 25 sources (see \S~\ref{clusterstructure}),
  we measure a protostellar density of $\sim 700$ pc$^{-3}$ and a
  minimum spanning tree $Q$-parameter of 0.82. {Although our
    measurement of the $Q$-parameter is likely limited in terms of
    sensitivity and extent,} the value is close to the expected value
  for a uniform volume density of sources.

\item All nine sources detected at more than one continuum wavelength
  have SEDs indicative of dust emission. The long wavelength emission
  toward SMA~1b and SMA~4 is well modeled by the additional presence
  of a HCHII region.  Thermal molecular line emission is detected
  towards six of the 1.3 mm continuum sources (SMA1b, SMA2, SMA4,
  SMA6, SMA15, and SMA18). From LTE modeling of CH$_3$CN ($J$=12-11)
  using CASSIS we find gas temperatures ranging from 95-373~K,
  CH$_3$CN column densities from (4-40)$\times 10^{16}$ cm$^{-2}$, and
  H$_2$ gas densities of (0.8-9)$\times 10^{8}$ cm$^{-3}$ (assuming a
  CH$_3$CN:H$_{2}$ abundance of $1\times 10^{-8}$).  The radial
  velocities of the hot cores, measured from CH$_3$CN, range from
  $-3.3$ to $-7.0$ km~s$^{-1}$, and the corresponding 1D velocity
  dispersion of {1.43 km~s$^{-1}$} implies a dynamical mass of
  {$410 \pm 260$ \msun}. This mass is compatible with the gas
  mass of $\sim 280$~\msun\/ based on single dish imaging, and
  demonstrates that hot core line emission can be an important probe
  of protocluster dynamics.

\item The dominant hot core SMA~1b shows a consistent spatial-velocity
  structure in a wide range of hot core molecular lines that is
  consistent with a disk undergoing Keplerian rotation and free-fall.
  The orientation of the disk is in excellent agreement with the major
  axis of the dust continuum emission, and is perpendicular to the
  large scale outflow axis. The outer radius of the disk is $\sim$800
  AU, and the enclosed mass is $\sim$10 - 30~\msun\/ (depending on the
  inclination angle).  {The radial distribution of HNCO and
    \hcccn\/ appears to differ from the rest of the molecules.}

\item Nine positions in the protocluster exhibit 229.7588 GHz Class I
  methanol maser emission, generally in close proximity to
  previously-identified 44 GHz or 24.9 GHz Class I masers.  The
  brightest position also exhibits 218.4400 GHz Class I methanol maser
  emission.  These are the first observations with sufficient angular
  resolution to directly establish maser activity in these transitions
  (by demonstrating brightness temperatures in excess of their
  excitation energies).



\end{itemize} 

\acknowledgments

Based on analysis carried out with the CDMS and JPL spectroscopic
databases and \url{splatalogue.net}.
This research has made use of NASA's Astrophysics Data System
Bibliographic Services.
This research made use of Astropy, a community-developed core Python
package for Astronomy \citep[http://www.astropy.org;][]{astropy}.  We
thank Fred Schwab for assistance in interpreting the equations of
\citet{Olnon75}.  We thank Ed Fomalont for his independent astrometric
analysis of the ALMA commissioning data from project 2010.2.99001.CSV
described in JIRA ticket CSV-2909.  We thank R. Friesen and
R. Indebetouw for providing useful comments on the manuscript.  We
thank the anonymous referee for a thorough report. ALMA is a
partnership of ESO (representing its member states), NSF (USA) and
NINS (Japan), together with NRC (Canada) and NSC and ASIAA (Taiwan),
in cooperation with the Republic of Chile. The Joint ALMA Observatory
is operated by ESO, AUI/NRAO and NAOJ.  We thank H. M\"uller for pointing
out an error in the quantum numbers listed for the 218~GHz maser
transition, which we corrected after publication.


\clearpage

\begin{deluxetable}{lccc} 
\rotate
\small
\tablewidth{0pc}
\tablecaption{Parameters of new observations\label{smaobs}}  
\tablecolumns{4}
\tablehead{\colhead{Parameter} & \colhead{SMA 1.3 mm} & \colhead{SMA 0.87 mm} & \colhead{VLA 6 cm}}
\startdata
Observing date         & 23 Feb 2010          &  18 Feb 2010        & 07 Jul 2011\\ %
On-source time         & 240 minutes     & 210 minutes  &  84 minutes \\
Project code           & 2009B-S036           & 2009B-S036          & 10C-186 \\
Antennas               & 7                    & 6                   & 27\\
Configuration          & very extended (VEX)  & very extended (VEX) & A\\
Projected baseline lengths  & 28 m -- 508 m & 26.5 m -- 508 m    & 0.53 km -- 36.63 km\\
J2000 phase center     & 17:20:55.0, -35:45:07.0 & 17:20:55.0, -35:45:07.0  & 17:20:53.33, -35:46:00.0\\
Primary beam FWHM      & 53\arcsec\/          & 34\arcsec\/    & 9.5\arcmin, 6.8\arcmin\/ \tablenotemark{a} \\
Synthesized beam\tablenotemark{b}  & 0.68\arcsec$\times$0.44\arcsec\/ (+5\arcdeg) & 0.55\arcsec$\times$0.26\arcsec\/ (+9\arcdeg) & 0.90\arcsec$\times$0.30\arcsec\/ (-2\arcdeg) \\
Synthesized beam (EXT+VEX) & 0.70\arcsec$\times$0.39\arcsec\/ (+5\arcdeg) & ... & ...\\
Synthesized beam (VEX-UV)\tablenotemark{c} & 0.66\arcsec$\times$0.36\arcsec\/ (+5\arcdeg) & 0.66\arcsec$\times$0.36\arcsec\/ (+5\arcdeg) & ...\\
Lower band center      & 218.85 GHz           &  335.54 GHz            & 5.06 GHz \\
Upper band center      & 230.83 GHz           &  347.54 GHz            & 7.16 GHz \\
Bandwidth              & 4 $\times$ 1.968 GHz & 4 $\times$ 1.968 GHz   & 2 $\times$ 1.024 GHz\\
Subbands               & 4 $\times$ 24        & 4 $\times$ 24          & 2 $\times$ 8\\
Polarization           & single linear        & single linear          & dual circular \\
Channel spacing        & 1.1 \kms             & 0.7 \kms               & 60 \kms, 43 \kms\/ \tablenotemark{a}\\
Continuum rms noise (EXT+VEX) \tablenotemark{d}    &  2.2 \mjb\ &  8.7 \mjb\/    & 18, 27 \mujb\/ \tablenotemark{a}\\
Spectral line rms noise \tablenotemark{d}         &  30 \mjb\    &   130 \mjb\/             & ...\\
Gain calibrator(s)     & J1733-1304, J1924-2914 & J1733-1304, J1924-2914  & J1717-3342 \\
Bandpass calibrator    & 3C273                  & 3C273                   & J1924-2914 \\ 
Flux calibrator(s)     & Titan, Ceres            & Ceres        & J1331+3030 \\
225 GHz zenith opacity & 0.07                    & 0.03-0.06    & n/a \\
Median wind speed      & 2 m/s                   & 5 m/s        & 3 m/s  \\
\enddata
%
\tablenotetext{a}{The first number is for the 5 GHz data; the second number is for the 7 GHz data.}
\tablenotetext{b}{This is the angular resolution of the 1.3~mm spectral line cubes, 0.87~mm continuum image, and 5 \& 7 GHz continuum images (convolved to the same resolution), respectively. The position angle in degrees east of north is given in parentheses.}
\tablenotetext{c}{The 1.3~mm image was generated from the uv spacings $>90$k$\lambda$, and the 0.87~mm image was generated with a uv-taper of 300~k$\lambda$ and then convolved to the same resolution as the 1.3~mm VEX-UV image.}
\tablenotetext{d}{Measured at the center of the primary beam.}
\end{deluxetable}

\begin{deluxetable}{lcccccccc}  
\rotate
\tablewidth{0pt} 
\tabletypesize{\footnotesize}
\tablecolumns{9}   
\tablecaption{Observed properties of the 1.3 mm continuum sources \label{contsources}}      
\tablehead{\colhead{} & \multicolumn{2}{c}{Fitted Position (J2000)} & \colhead{Peak intensity\tablenotemark{b}} & \colhead{Peak} & 
           \colhead{Fitted flux}  & \colhead{Fitted size (Pos. angle E of N)\tablenotemark{c}} & Fitted size & \colhead{T$_{\rm brightness}$} \\
  \colhead{Source\tablenotemark{a,e}}  & \colhead{$\alpha$ ($^{\rm h}~~^{\rm m}~~^{\rm s}$)} & \colhead{$\delta$ ($^{\circ}~~{'}~~{''}$)} 
        &  \colhead{(\jb)}  & \colhead{S/N}  & \colhead{density (Jy)}    & \colhead{${''} \times {''}$ (\arcdeg)} & \colhead{(AU)} & \colhead{(K)} }  
\startdata
SMA 1a   & 17:20:55.137 & -35:45:05.76 & 0.0866  $\pm$0.0022 &  39   & 0.225$\pm$0.020 & 0.99$\pm$0.05 $\times$ 0.52$\pm$0.10 (+35$\pm$5)  & 1300$\times$680 & 10$\pm$2\\ 
SMA 1b+d & 17:20:55.192 & -35:45:03.93 & 0.3257  $\pm$0.0022 & 148   & 1.059$\pm$0.076 & 1.02$\pm$0.05 $\times$ 0.74$\pm$0.06 (+128$\pm$9) & 1300$\times$960 & 32$\pm$4\\ 
SMA 1c   & 17:20:55.267 & -35:45:02.89 & 0.1384  $\pm$0.0022 &  63   & 0.335$\pm$0.024 & 0.74$\pm$0.05 $\times$ 0.61$\pm$0.07 (+141$\pm$25)& 960$\times$790 & 17$\pm$3\\
SMA 2    & 17:20:54.870 & -35:45:06.40 & 0.1039  $\pm$0.0022 &  47   & 0.169$\pm$0.007 & 0.58$\pm$0.03 $\times$ 0.27$\pm$0.07 (+138$\pm$6) & 750$\times$350 & 24$\pm$6\\
SMA 3    & 17:20:55.003 & -35:45:07.40 & 0.0331  $\pm$0.0022 &  15   & 0.138$\pm$0.016 & 1.06$\pm$0.06 $\times$ 0.93$\pm$0.09 (+10$\pm$25) & 1400$\times$1200 & 3.2$\pm$0.5\\ 
SMA 4    & 17:20:54.627 & -35:45:08.73 & 0.0425  $\pm$0.0022 &  19   & 0.059$\pm$0.004 & $<0.5$ & $<650$ & $>5.4$ \\ 
SMA 5    & 17:20:55.043 & -35:45:01.53 & 0.0433  $\pm$0.0022 &  20   & 0.053$\pm$0.002 & 0.68$\pm$0.01 $\times$ 0.51$\pm$0.03 (+6$\pm$1)   & 890$\times$660 & 4.9$\pm$0.3\\
SMA 6    & 17:20:54.590 & -35:45:17.40 & 0.1812  $\pm$0.0022 &  82   & 0.285$\pm$0.016 & $<0.5$ & $<650$ & $>26$ \\ 
SMA 7    & undetected   & ...          &  ...     & ...	 &       ...       & ...   & ...           \\	  	  
SMA 8    & 17:20:53.782 & -35:44:46.16 & 0.0297  $\pm$0.0034 &   8.7 & 0.046$\pm$0.003 & $<0.5$ & $<650$ & $>4.4$ \\ 
SMA 9    & 17:20:54.196 & -35:45:41.45 & 0.1070  $\pm$0.0054 &  20   & 0.139$\pm$0.006 & $<0.5$ & $<650$ & $>13$ \\ 
SMA 10   & 17:20:54.283 & -35:45:09.38 & 0.0115  $\pm$0.0022 &   5.2 & 0.025$\pm$0.003 & $<0.5$ & $<650$ & $>2.2$ \\
SMA 11   & 17:20:54.751 & -35:45:20.18 & 0.0197 $\pm$0.0022 &   9.0 & 0.043$\pm$0.004 & 0.72$\pm$0.07 $\times$ 0.27$\pm$0.15 (+76$\pm$25) & 940$\times$350 & 5.0$\pm$2.5\\                 
SMA 12   & 17:20:54.849 & -35:45:33.51 & 0.0312 $\pm$0.0034 &   9.2 & 0.039$\pm$0.004 & $<0.5$ & $<650$ & $>3.5$\\ 
SMA 13   & 17:20:54.900 & -35:45:16.41 & 0.0425 $\pm$0.0022 &  19   & 0.049$\pm$0.004 & $<0.5$ & $<650$ & $>4.4$ \\ 
SMA 14   & 17:20:54.962 & -35:45:08.92 & 0.0130	$\pm$0.0022 &   5.9 & 0.022$\pm$0.002 & 0.49$\pm$0.09 $\times$ 0.39$\pm$0.11 (+104$\pm$25) & 640$\times$510 & 2.7$\pm$0.9  \\
SMA 15   & 17:20:55.504 & -35:45:10.96 & 0.0469 $\pm$0.0022 &  21   & 0.063$\pm$0.003 & $<0.5$ & $<650$ & $>5.8$ \\ 
SMA 16   & 17:20:55.503 & -35:44:55.90 & 0.0179 $\pm$0.0022 &   8.1 & 0.031$\pm$0.004 & $<0.5$ & $<650$ & $>2.8$ \\ 
SMA 17   & 17:20:55.590 & -35:44:56.89 & 0.0145 $\pm$0.0022 &   6.6 & 0.021$\pm$0.003 & $<0.5$ & $<650$ & $>1.9$ \\ 
SMA 18\tablenotemark{d} 
         & 17:20:55.645 & -35:45:32.61 & 0.0433 $\pm$0.0034 &  13   &  0.043$\pm$0.003 & $<0.5$ & $<650$ & $>3.9$ \\ 
SMA 19   & 17:20:55.854 & -35:45:26.98 & 0.0179 $\pm$0.0025 &   7.2 & 0.024$\pm$0.004 & $<0.5$  & $<650$ & $>2.1$    \\ 
SMA 20   & 17:20:56.061 & -35:45:32.66 & 0.0320 $\pm$0.0034 &   9.4 & 0.032$\pm$0.003 & $<0.5$  & $<650$ & $>2.4$   \\ 
SMA 21   & 17:20:56.298 & -35:45:27.04 & 0.0362 $\pm$0.0034 &  11   & 0.052$\pm$0.004 & $<0.5$  & $<650$ & $>4.7$\\
SMA 22   & 17:20:56.571 & -35:45:17.01 & 0.0263 $\pm$0.0025 &  11 & 0.042$\pm$0.003 & 0.58$\pm$0.05 $\times$ 0.31$\pm$0.10 (+2$\pm$4) & 750$\times$400 & 5.5$\pm$1.9\\
SMA 23   & 17:20:56.629 & -35:45:09.84 & 0.0167 $\pm$0.0025 &   6.7 & 0.017$\pm$0.002 & $<0.5$  & $<650$ & $>1.5$        \\
\enddata 
\tablenotetext{a}{SMA~1 through 7 were previously detected in \citet{Brogan09}}
\tablenotetext{b}{The intensity of the peak pixel; uncertainties correspond to the image rms in the annular region in which the source is located (see \S~\ref{annulus}).}
\tablenotetext{c}{Deconvolved from the beam}
\tablenotetext{d}{Position consistent with SM2 from \citet{Sandell00}}
\tablenotetext{e}{Quantities in columns labeled ``Fitted'' were obtained 
from the CASA imfit task}
\end{deluxetable}				 			

\clearpage

\begin{table} 
\caption{Observed properties of the 6 centimeter continuum sources \label{cmcontsources}}  
\begin{tabular}{lcccc}    
\tableline
\tableline
        & \multicolumn{2}{c}{Fitted Position (J2000)\tablenotemark{a}}  & \multicolumn{2}{c}{Peak intensity (\mjb)}\tablenotemark{b} \\
Source  & $\alpha$ ($^{\rm h}~~^{\rm m}~~^{\rm s}$)   & $\delta$ ($^{\circ}~~{'}~~{''}$) 
                                                        &  5.06 GHz & 7.16 GHz \\ 
\tableline
SMA 1b                      & 17 20 55.192 & -35 45 03.83 &  0.327$\pm$0.014 & 0.373$\pm$0.019 \\
SMA 4                       & 17 20 54.619 & -35 45 08.57 &  0.157$\pm$0.014 & 0.235$\pm$0.026 \\
H$_2$O-C4\tablenotemark{c} & 17 20 54.149 & -35 45 13.70 &  0.110$\pm$0.011 & 0.091$\pm$0.019 \\
VLA 2\tablenotemark{b}      & 17 20 54.871 & -35 45 36.49 &  0.102$\pm$0.014 & 0.133$\pm$0.041 \\
\tableline
\tablenotetext{a}{Obtained from a 2D Gaussian fit to the image}
\tablenotetext{b}{The intensity of the peak pixel; measured from the image corrected for primary beam response and convolved to 0.9\arcsec $\times$ 0.3\arcsec\/ beam}
\tablenotetext{c}{newly-detected cm sources}
\end{tabular}
\end{table}				 			

\begin{deluxetable}{lcc}     
\tablewidth{0pt} 
\tablecolumns{3}   
\tablecaption{Revised positions of the 1.3 and 0.7 centimeter continuum sources \label{cmpositions}}  
\tablehead{ & \multicolumn{2}{c}{Fitted Position (J2000)\tablenotemark{a}} \\
      \colhead{Source}  & \colhead{$\alpha$ ($^{\rm h}~~^{\rm m}~~^{\rm s}$)} & \colhead{$\delta$ ($^{\circ}~~{'}~~{''}$)}} 
\startdata
\cutinhead{1.3 Centimeter}
SMA 1b  & 17 20 55.185 & -35 45 03.98 \\ 
SMA 1c  & 17 20 55.259 & -35 45 02.88 \\  
SMA 1d  & 17 20 55.222 & -35 45 04.08 \\ 
SMA 4   & 17 20 54.616 & -35 45 08.66 \\
\cutinhead{0.7 Centimeter}
SMA 1a  & 17 20 55.150 & -35 45 05.71 \\
SMA 1b  & 17 20 55.190 & -35 45 03.95 \\
SMA 1c  & 17 20 55.255 & -35 45 02.91 \\
SMA 1d  & 17 20 55.223 & -35 45 04.12 \\
VLA 3   & 17 20 54.992 & -35 45 06.92 \\
SMA 4   & 17 20 54.618 & -35 45 08.71 \\
SMA 6   & 17 20 54.591 & -35 45 17.39 \\
\enddata
\tablenotetext{a}{These positions were extracted from {Gaussian fits to} 
images corrected for the position error of the phase calibrator J1720-358 
in the VLA catalog, 
as described in \S~\ref{obs}. The uncertainties on the absolute fitted positions
are $\leq 0.03$\arcsec.}
\end{deluxetable}

\begin{deluxetable}{lccccccc} 
\rotate
\tablewidth{0pt} 
\tablecolumns{8}
\tabletypesize{\small}
\tablecaption{Observed properties of the 0.87 mm continuum sources \label{contsourcesBand7}}  
\tablehead{ & \multicolumn{2}{c}{Fitted Position (J2000)}
        & \colhead{Peak intensity\tablenotemark{a}} & \colhead{Fitted flux\tablenotemark{b}}  & \colhead{Fitted size (Pos. angle E of N)\tablenotemark{c}} & \colhead{Fitted size} & \colhead{T$_{\rm brightness}$}\\
\colhead{Source\tablenotemark{d}}  & \colhead{$\alpha$ ($^{\rm h}~~^{\rm m}~~^{\rm s}$)}   & \colhead{$\delta$ ($^{\circ}~~{'}~~{''}$)}  
        &  \colhead{(\jb)}   & \colhead{density (Jy)}  & \colhead{${''} \times {''}$ (\arcdeg)} & (AU) & \colhead{(K)}} 
\startdata
SMA 1c   & 17:20:55.269 & -35:45:02.88 & 0.296$\pm$0.014 &  0.63$\pm$0.03  & 0.63$\pm$0.03 $\times$ 0.42$\pm$0.05 (+113$\pm$10) & 820$\times$550 & 24$\pm$3\\
SMA 1b+d & 17:20:55.194 & -35:45:03.98 & 0.801$\pm$0.014 &  1.67$\pm$0.08  & 0.62$\pm$0.03 $\times$ 0.36$\pm$0.06 (+129$\pm$8) & 810$\times$470 & 76$\pm$14\\
SMA 2    & 17:20:54.871 & -35:45:06.43 & 0.230$\pm$0.014 &  0.27$\pm$0.02  & 0.64$\pm$0.03 $\times$ 0.43$\pm$0.04 (+164$\pm$2) & 830$\times$560 & 9.9$\pm$1.2\\
SMA 4    & 17:20:54.626 & -35:45:08.73 & 0.125$\pm$0.015 &  0.13$\pm$0.02  & 0.65$\pm$0.02 $\times$ 0.38$\pm$0.04 (+178$\pm$2) & 850$\times$490 & 5.4$\pm$0.7\\
SMA 5    & 17:20:55.048 & -35:45:01.42 & 0.061$\pm$0.016 &  0.06$\pm$0.02  & $<0.5$ & $<650$ & $>3.5$\\
SMA 6    & 17:20:54.595 & -35:45:17.40 & 0.461$\pm$0.020 &  0.66$\pm$0.04  & $<0.5$ & $<650$ & $>41$\\
SMA 13   & 17:20:54.883 & -35:45:16.32 & 0.107$\pm$0.018 &  0.28$\pm$0.03  & $<0.5$ & $<650$ & $>17$\\
SMA 15   & 17:20:55.511 & -35:45:11.07 & 0.107$\pm$0.016 &  0.19$\pm$0.04  & $<0.5$ & $<650$ & $>12$\\ 
\enddata
\tablenotetext{a}{The intensity of the peak pixel.}
\tablenotetext{b}{Due to the difference in uv coverage, these values should be considered lower limits when compared 
to the 1.3~mm flux densities in Table~\ref{contsources}, 
except for SMA~13 and 15.  The uncertainties do not include the overall calibration uncertainty of 20\%}
\tablenotetext{c}{Deconvolved from the beam}
\tablenotetext{d}{Quantities in columns labeled ``Fitted'' were obtained 
from the CASA imfit task}
\end{deluxetable}

\begin{deluxetable}{lcccccccccc}
\tablecaption{Properties of spectral lines shown in Figure~\ref{moments}. \label{linelist}}
\rotate
\tablecolumns{11}
\tablewidth{0pt}
\setlength{\tabcolsep}{0.06in} 
\tabletypesize{\footnotesize}
\tablehead{\colhead{Species} & \colhead{Transition} & \colhead{Frequency}  & \colhead{$E_{lower}$} & \colhead{} &  
           \multicolumn{6}{c}{Integrated Intensity (Jy beam$^{-1}$*\kms\/)\tablenotemark{c}} \\
            & & \colhead{(GHz)}      & \colhead{(K)}  &  
           \colhead{Catalog\tablenotemark{a,b}} & \colhead{SMA1b} & \colhead{SMA2} & \colhead{SMA4} & 
           \colhead{SMA6} & \colhead{SMA15} & \colhead{SMA18}}
\startdata
\hcccn\                                      & J=24--23 & 218.32479 & 120.5 & JPL   & 7.49 $\pm$ 0.22  & $<$0.65      & 2.48 $\pm$ 0.22  & 1.52 $\pm$ 0.38  & $<$0.74  &  2.90 $\pm$ 0.41 \\
OCS                                          & J=18--17 & 218.90336 & 89.3 & CDMS   & 4.71 $\pm$ 0.21  & 1.38 $\pm$ 0.21  & 2.64 $\pm$ 0.21  & 2.03 $\pm$ 0.28  & 0.6 $\pm$ 0.2  &  2.36 $\pm$ 0.36  \\
$^{13}$CS                                     & J=5--4   & 231.22069 & 22.2 & CDMS   & 2.06 $\pm$ 0.30  & 1.22 $\pm$ 0.30  & 2.06 $\pm$ 0.30  & 1.02 $\pm$ 0.29  & $<$0.70  &  $<$1.54  \\
\methanol\/ ($E$)               & 8$_{0,8}$--7$_{1,6}$     & 220.07849 & 86.1 & CDMS  & 5.45 $\pm$ 0.33  & 3.16 $\pm$ 0.33  & 4.48 $\pm$ 0.33  & 2.04 $\pm$ 0.32  & $<$0.79  &  $<$1.27 \\
\methanol\/ ($E$)              & 20$_{1,19}$--20$_{0,20}$ & 217.88639 & 497.9 & CDMS  & 3.68 $\pm$ 0.22  & 1.07 $\pm$ 0.22  & 2.63 $\pm$ 0.22  & $<$0.86            & $<$0.84  &  $<$1.31  \\
\methanol\/ v$_t$=1 ($A$)           & 6$_{1,5}$--7$_{2,6}$ & 217.29920 & 363.5 & CDMS & 4.40 $\pm$ 0.28  & 1.61 $\pm$ 0.28  & 3.47 $\pm$ 0.28  & $<$0.87            & $<$0.86  &  $<$1.72 \\
\methylcyanide\                            & J=12--11, K=3 & 220.70902 & 122.6 & JPL & 6.08 $\pm$ 0.27  & 2.56 $\pm$ 0.27  & 3.29 $\pm$ 0.27  & 1.87 $\pm$ 0.35 & $<$0.80  &  2.63 $\pm$ 0.39 \\
\methylcyanide\                            & J=12--11, K=7 & 220.53932 & 408.0 & JPL & 2.63 $\pm$ 0.22  & $<$0.65      & $<$0.65      & $<$0.74           & $<$0.49  &  $<$1.08 \\
\ethylcyanide\                  & 25$_{2,24}$--24$_{2,23}$ & 220.66092 & 132.4 & CDMS  & 4.34 $\pm$ 0.29  & $<$0.88      & $<$0.88      & $<$0.71           & $<$0.66  &  $<$1.18 \\
CH$_3$OCHO ($A$)\tablenotemark{d} & 20$_{1,20}$--19$_{1,19}$ & 216.96590 & 101.1 & JPL & 5.75 $\pm$ 0.28  & 4.34 $\pm$ 0.28  & 2.53 $\pm$ 0.28  & 2.03 $\pm$ 0.35  & $<$0.99  & $<$1.52  \\
\dimethylether\                   & 17$_{2,15}$--16$_{3,14}$ & 230.23376 & 136.6 & JPL & 3.19 $\pm$ 0.23  & 1.23 $\pm$ 0.23  & 1.06 $\pm$ 0.23  & $<$0.72            & $<$0.63  & $<$1.25  \\
HNCO                                & 10$_{1,10}$--9$_{1,9}$ & 218.98101 & 90.6 & CDMS & 5.28 $\pm$ 0.23  & 2.13 $\pm$ 0.23  & 2.07 $\pm$ 0.23  & 0.85 $\pm$ 0.27  & $<$0.89  & $<$1.44 \\
\enddata
\tablenotetext{a}{CDMS =
\url{http://www.astro.uni-koeln.de/cgi-bin/cdmssearch}}
\tablenotetext{b}{JPL =
\url{http://spec.jpl.nasa.gov/ftp/pub/catalog/catform.html}}
\tablenotetext{c}{The integrated intensity was measured from moment zero images; upper limits are 3$\sigma$.}
\tablenotetext{d}{This transition is blended with three other
CH$_3$OCHO transitions with the same line intensity and $E_{\rm
lower}$.  The velocity separations from this reference transition are
+1.57, -0.48, and -2.10~\kms, which is less than half of the total
velocity gradient in SMA~1b.}
\end{deluxetable} 

\begin{deluxetable}{lccccccc}    
\tablecolumns{8} 
\tablewidth{0pc}
\tablecaption{Best fit CASSIS LTE models to the \methylcyanide\/ spectra \label{temperatures}}   
\tabletypesize{\footnotesize}
\tablehead{\colhead{} & \colhead{Temperature} & \multicolumn{2}{c}{Diameter} & \colhead{Column density} & \colhead{Linewidth} & 
             \colhead{LSR velocity} & \colhead{H$_2$ Density\tablenotemark{a}}\\ 
           \colhead{Source} & \colhead{(K)}   & \colhead{(\arcsec\/)}   & \colhead{(AU)} & \colhead{(\percms\/)} & 
              \colhead{(\kms)} & \colhead{(\kms)}  & \colhead{(\percmc)}  }
\startdata
\cutinhead{One-component models} 
SMA 1b &     143$\pm$7  & 0.38$\pm$0.01 & 490 & (3.94$\pm$0.34)E17 & 4.72$\pm$0.01 & -3.29$\pm$0.01 & 8E9\\  
SMA 2  &     140$\pm$6  & 0.30$\pm$0.01 & 390 & (4.26$\pm$0.72)E16 & 4.40$\pm$0.07 & -7.01$\pm$0.02 & 1E9\\  
SMA 4  &     208$\pm$2  & 0.26$\pm$0.01 & 340 & (6.26$\pm$0.20)E16 & 8.79$\pm$0.04  & -6.80$\pm$0.02 & 2E9\\  
SMA 6  &      95$\pm$3  & 0.25$\pm$0.01 & 330 &  (9.39$\pm$0.32)E16 & 4.26$\pm$0.15 & -4.91$\pm$0.10 & 3E9\\  
SMA 15 &      72$\pm$15 & 0.22$\pm$0.03 & 290 & (3.76$\pm$0.46)E16 & 3.76$\pm$0.46 & -6.36$\pm$0.37 & 1E9\\  
SMA 18 &     139$\pm$10 & 0.25$\pm$0.02 & 330 & (4.77$\pm$0.51)E16 & 5.71$\pm$0.36 & -4.94$\pm$0.16 & 1E9\\  
\cutinhead{Two-component models} 
SMA 1b comp. 1 & 80$\pm$5   & 0.40$\pm$0.02 & 520 & (4.51$\pm$0.31)E17 & 3.21$\pm$0.15 & -3.24$\pm$0.17 & 9E9 \\   
SMA 1b comp. 2 & 307$\pm$40 & 0.24$\pm$0.02 & 310 & (2.89$\pm$0.27)E17 & 6.63$\pm$0.52 & -3.35$\pm$0.17 & 9E9\\   
SMA 4 comp. 1 &  135$\pm$5  & 0.30$\pm$0.01 & 390 & (3.06$\pm$0.25)E16 & 4.78$\pm$0.66 & -8.25$\pm$0.07 & 8E8 \\  
SMA 4 comp. 2 &  280$\pm$35 & 0.22$\pm$0.02 & 290 & (3.31$\pm$0.51)E16 & 6.85$\pm$0.38 & -2.67$\pm$0.19 & 1E9\\  
\enddata
\tablenotetext{a}{Derived from the CH$_{3}$CN column density assuming spherical geometry and a \methylcyanide:H$_2$ abundance of $\sim 1 \times 10^{-8}$.}
\end{deluxetable}

\begin{table} 
\footnotesize
\caption{Observed and derived parameters of the 218 and 229~GHz \methanol\/ maser emission \label{masertable}}   
\begin{tabular}{lccccccc}   
  \tableline
  & \multicolumn{2}{c}{Fitted Position (J2000)\tablenotemark{a}}  &  Peak intensity & Flux density & Deconvolved & $T_{\rm B}$ & $v_{\rm LSR}$\\
  Component\tablenotemark{b}   & $\alpha$ ($^{\rm h}~~^{\rm m}~~^{\rm s}$)   & $\delta$ ($^{\circ}~~{'}~~{''}$) & (Jy/beam)  & (Jy)
  & size (\arcsec $\times$ \arcsec) & (K)\tablenotemark{c} & (\kms) \\ 
  \tableline
  \tableline 
  \multicolumn{7}{c}{229.75881 GHz (8$_{\rm -1,8}$-7$_{\rm 0,7}$)~E transition}\\
  1 (8)  & 17 20 54.327 & -35 45 22.31 & 0.76 & $1.51 \pm 0.12$ & $<0.5$             & 145    & $-7.56 \pm 0.06$\\ 
  2 (9)  & 17 20 54.392 & -35 45 22.68 & 0.57 & $1.91 \pm 0.13$ & $<0.5$             & 180    & $-8.66 \pm 0.13$\\ 
  3 (11) & 17 20 54.447 & -35 45 22.76 & 1.46 & $3.19 \pm 0.15$ & $0.67 \times 0.23$ & 480    & $-7.65 \pm 0.07$\\ 
  4      & 17 20 54.589 & -35 45 10.33 & 0.79 & $2.18 \pm 0.22$ & $0.80 \times 0.42$ & 150    & $-2.66 \pm 0.06$\\
  5 (22) & 17 20 54.925 & -35 45 13.70 & 1.25 & $1.45 \pm 0.18$ & $<0.5$             & $>140$ & $-5.56 \pm 0.06$\\ 
  6 (25) & 17 20 54.952 & -35 45 13.99 & 0.84 & $1.98 \pm 0.11$ & $0.82 \times 0.15$ & 380    & $-5.15 \pm 0.07$\\ 
  7\tablenotemark{d}
  & 17 20 55.048 & -35 45 14.89 & 0.76 & $0.84 \pm 0.06$ & $0.39 \times 0.12$ & 420    & $-5.73 \pm 0.08$\\ 
  8\tablenotemark{d}
  & 17 20 55.024 & -35 45 15.32 & 0.60 & $1.60 \pm 0.07$ & $0.82 \times 0.42$ & 110    & $-5.15 \pm 0.06$\\ 
  9\tablenotemark{d} (55) 
  & 17 20 55.670 & -35 45 00.38 & 6.40 & $8.80 \pm 0.16$ & $0.44 \times 0.15$ & 3100   & $-3.03 \pm 0.01$\\ 
  \tableline
  \multicolumn{7}{c}{218.44005 GHz ($4_{2,2}$-$3_{1,2}$)~E transition}\\
  9 (55) & 17 20 55.671 & -35 45 00.34 & 2.28 & $3.09 \pm 0.12$ & $<0.5$             & 270    & $-3.07 \pm 0.03$\\ 
\tableline
\end{tabular}
\tablenotetext{a}{The fitted position of the component in the channel of peak emission.}
\tablenotetext{b}{A number in parentheses indicates that the corresponding 44~GHz \methanol\/ maser component 
from Table~4 of \citet{Brogan09} is within 0.2\arcsec\/ of this position.}
\tablenotetext{c}{The brightness temperature ($T_{\rm B}$) is computed from the integrated flux density and the deconvolved
size.  When the deconvolved size is an upper limit, the beam size is used (0.71\arcsec\/$\times$0.37\arcsec\/ 
for the 218.4400~GHz line, and 0.67\arcsec\/$\times$0.36\arcsec\/ for the 229.7588 GHz line).}
\tablenotetext{d}{This component is within 0.3\arcsec\/ of a 24.9~GHz \methanol\/ maser \citep{Beuther05}.}
\end{table}

\begin{table} 
\caption{Gas properties derived from the dust emission \label{dustmass}}   
\begin{tabular}{lccc}   
\tableline
       & Temperature range\tablenotemark{a} & Mass range\tablenotemark{b} & Column density range\tablenotemark{b}\\ 
Source &    (K) & (\msun)                     & (10$^{24}$ \percms) \\ 
\tableline
\tableline
SMA 1a   & 20 --  50 &  2.2 --  15   &  4.8 -- 33   \\
SMA 1b   & 143$\pm$7 & 4.3$\pm$1.1   &  1.0$\pm$0.2   \\ 
SMA 1c   & 20 --  50 &  3.5 --  31   &  9.6 -- 92   \\
SMA 1d   & 20 --  50 &  0.4 -- 2.2   &  1.9 -- 11   \\
SMA 2    & 140$\pm$6 & 0.71$\pm$0.18 & 8.1$\pm$2.7   \\ 
SMA 3    & 20 --  50 &   1.2 --  6.5 &  1.5 -- 8.1    \\
SMA 4    & 208$\pm$2 & 0.15$\pm$0.03 & 1.1$\pm$0.2   \\ 
SMA 5    & 20 --  50 &   0.5 --  2.5 &  1.8 -- 8.5   \\
SMA 6    & 95$\pm$3  & 2.0$\pm$0.5   & 1.4$\pm$0.3   \\ 
SMA 8    & 20 --  50 &   0.4 --  2.2 &  2.1 -- 10    \\
SMA 9    & 20 --  50 &   1.4 --  11  &  7.3 -- 46    \\
SMA 10   & 20 --  50 &   0.2 --  1.1 &  1.1 -- 5.5    \\
SMA 11   & 20 --  50 &   0.4 --  2.2 &  1.5 -- 16    \\
SMA 12   & 20 --  50 &   0.4 --  1.8 &  1.7 -- 8.8    \\
SMA 13   & 20 --  50 &   0.5 --  2.4 &  2.3 -- 11    \\
SMA 14   & 20 --  50 &  0.2 --  1.0  &  1.0 -- 7.3   \\
SMA 15   & 72$\pm$15 & 0.50$\pm$0.16   & 3.6$\pm$0.8    \\ 
SMA 16   & 20 --  50 &   0.3 --  1.5   &  1.4 -- 6.9    \\
SMA 17   & 20 --  50 &  0.2 --  1.0    &  1.3 -- 6.6  \\
SMA 18   & 139$\pm$10 &  0.17$\pm$0.04 & 1.2$\pm$0.3    \\ 
SMA 19   & 20 --  50 &  0.2 --  1.1    &  1.0 -- 5.4  \\
SMA 20   & 20 --  50 &  0.3 --  1.5   &  1.4 -- 7.1   \\
SMA 21   & 20 --  50 &  0.5 --  2.5   &  2.4 -- 12   \\
SMA 22   & 20 --  50 &  0.4 --  2.2   &  2.1 -- 16   \\
SMA 23   & 20 --  50 &  0.2 --  0.8   &  0.7 -- 3.6   \\
\end{tabular}
\tablenotetext{a}{For SMA~1b, 2, 4, 6, 15 and 18, we use the fitted gas temperature from
Table~\ref{temperatures}.}
\tablenotetext{b}{In calculating the mass and column density uncertainties and ranges, 
  we ran Monte-Carlo simulations using the uncertainties in the fitted temperature,
  flux density, and size.  The uncertainty in the flux density also includes
  20\% of the fitted flux density in order to account for the overall calibration
  accuracy (see \S~\ref{obs}). For sources with ranges, the smaller values of mass 
  and column density correspond to the higher temperature values.}

\end{table}

\clearpage

\clearpage

\clearpage

\clearpage

\clearpage

\clearpage

\clearpage

\clearpage

\clearpage

\clearpage

\clearpage

\clearpage

\end{document}